\documentclass[twocolumn,prb,aps,showpacs,amsmath,amssymb]{revtex4}

\usepackage{graphicx}
\usepackage{dcolumn}
\usepackage{bm}
\usepackage{times}

\renewcommand{\vec}[1]{{\bf#1}}

\begin{document}

\title{Interplay between quantum dissipation and an in-plane magnetic field in the spin ratchet effect}

\author{Sergey Smirnov,$^1$ Dario Bercioux,$^{2}$ Milena Grifoni,$^1$ and Klaus Richter$^1$}
\affiliation{$^1$Institut f\"ur Theoretische Physik, Universit\"at Regensburg, D-93040 Regensburg, Germany\\
  $^2$Physikalisches Institut and Freiburg Institute for Advanced Studies, Universit\"at Freiburg, D-79104 Freiburg,
  Germany}

\date{\today}

\begin{abstract}
We investigate the existence of the pure spin ratchet effect in a dissipative quasi-one-dimensional system with Rashba
spin-orbit interaction. The system is additionally placed into a transverse uniform stationary in-plane magnetic field. It
is shown that the effect exists at low temperatures and pure spin currents can be generated by applying an unbiased ac
driving to the system. An {\it analytical} expression for the ratchet spin current is derived. From this expression it
follows that the spin ratchet effect appears as a result of the simultaneous presence of the spin-orbit interaction,
coupling between the orbital degrees of freedom and spatial asymmetry. In this paper we consider the case of a broken
spatial symmetry by virtue of asymmetric periodic potentials. It turns out that an external magnetic field does not have
any impact on the existence of the spin ratchet effect, but it influences its efficiency enhancing or reducing the
magnitude of the spin current.
\end{abstract}


\maketitle

\section{Introduction}\label{intro}
It is well known that a directed stationary flow of particles in a system can be created by unbiased external forces. In
general this possibility arises when the system is not invariant under reflections in real space. This fact is mainly
independent of the mechanics which underpins the particle motion, classical or quantum. However, the microscopic origin of
this effect, conventionally called the ratchet effect, is different in the classical and quantum case. One principle
source of that difference is quantum mechanical tunnelling which does not have analogs in the classical mechanics.
Correspondingly, one usually distinguishes between classical and quantum ratchet effects. In this paper we concentrate on
the latter one in a dissipative system. Such dissipative ratchet systems act as Brownian motors \cite{Astumian,Haenggi}
turning Brownian into directed motion. The existence of the ratchet effect in a quantum dissipative one-dimensional (1D)
system which lacks the spatial symmetry has been first theoretically predicted in Ref.~\onlinecite{Reimann}. Later, within
a tight-binding model where the lowest bands are narrow, it has been disclosed that a ratchet state of the particle
transport can only be achieved when at least the two lowest Bloch bands contribute to transport \cite{Grifoni}. To obtain
the ratchet effect in systems with weak periodic potentials at least two harmonics of the potential should enter the
dynamical equations \cite{Peguiron}. Rectification can also take place in a single-band tight-binding model where the
spatial asymmetry is concealed from the electron dynamics. One way to achieve this is to use unbiased external forces with
harmonic mixing\cite{Goychuk}.

Coherent charge ratchets based on molecular wires with an asymmetric level structure of the orbital energies were proposed
in Ref.~\onlinecite{Lehmann}. In this case weak dissipation originates from a weak coupling between the wire edges and
leads which represent fermionic reservoirs. In contrast to the systems described above in this system there is no
dissipation in the wire. The ratchet effect is a result of the dissipative coupling of the wire to fermionic baths.

In a different branch of condensed matter a new research field has emerged during the last decade, namely spintronics,
where one tries to make use of the spin degree of freedom of a particle instead of only the charge one. One essential
difference between spin and charge is that a particle can have more than one spin state while it has only one charge
state. In the context of transport it is important that the spin state of a particle can strongly depend on the transport
conditions, in particular on the transport direction, as it happens for example in systems with spin-orbit interaction.
This fact has founded a new arena for different spin devices used to store, transform and transfer miscellaneous
information. The possibility to transfer the spin separately from charge plays an important role. This can be implemented
by so-called pure spin currents, that is spin currents which are not accompanied by charge currents. Thus the generation
of such currents has been extensively discussed. Among different mechanisms of spin-orbit interaction Rashba spin-orbit
interaction (RSOI) \cite{Rashba} plays a distinguished role because it provides an opportunity to control the spin-orbit
coupling strength by an external electric field. The change in the band structure spawned by the spin-orbit interaction
leads to one of the most remarkable effects in spintronics, the intrinsic spin-Hall effect, first predicted by Murakami
{\it et al.}\cite{Murakami} for hole-doped semiconductors with the spin-orbit interaction of the effective Luttinger model
for holes and later by Sinova {\it et al.}\cite{Sinova} in a high-mobility two-dimensional electron gas (2DEG) with RSOI.
The spin current which results from the intrinsic spin-Hall effect is pure and its experimental detection was discussed,
{\it e.g.}, by Wunderlich {\it et al.}\cite{Wunderlich} Another kind of spin-Hall effect, the extrinsic spin-Hall effect,
is a result of the spin-orbit interaction as well. The spin currents related to the extrinsic spin-Hall effect are also
pure. Such pure spin currents were experimentally detected through optical measuring of electron spin accumulation at the
edges of the samples\cite{Kato} and through the reciprocal spin-Hall effect \cite{Hankiewicz,Hankiewicz_1} in
Ref.~\onlinecite{Valenzuela}. Another approach to create pure spin currents is to use polarized light. For example in
noncentrosymmetric semiconductors one-photon absorption of linearly polarized light induces pure spin currents
\cite{Bhat}. The pure spin current response to linearly and circularly polarized light irradiation, exciting electrons
from valence bands into the conduction bands, was studied by Li {\it et al.}\cite{Li} and by Zhou {\it et al.}\cite{Zhou}
for 2DEGs with RSOI. An alternative technique of getting pure spin currents is quantum spin pumping. The idea of quantum
spin pumping comes from the general idea of electron pumping \cite{Altshuler}. Electron pumping assumes that in a given
system any voltage bias is absent and the particle flow is a result of a cyclic variation of at least two parameters of
this system. When the electron spin is involved due to some mechanisms, various quantum spin pumps emerge. For example
spin pumps based on electronic interactions \cite{Sharma}, magnetic barriers \cite{Benjamin}, carbon nanotubes \cite{Wei}
have been discussed. A spin pump based on a quantum dot was experimentally implemented by Watson {\it et al.}\cite{Watson}
The pure spin current generation using the spin ratchet effect in coherent and dissipative systems with RSOI was
investigated in Refs.~\onlinecite{Scheid} and \onlinecite{Smirnov}, respectively. The spin ratchet effect in the presence
of a non-uniform static magnetic field without spin-orbit interaction, the Zeeman ratchet effect, was studied in
Ref.~\onlinecite{Scheid_1} for coherent quantum wires formed in a 2DEG. However, the spin ratchet effect in a dissipative
system in an external magnetic field has not been considered up to now.

In this paper we extend the results of Ref.~\onlinecite{Smirnov} to include a transverse in-plane uniform stationary
magnetic field. Specifically, we consider non-interacting electrons in a quantum wire formed by a harmonic transverse
confinement in a 2DEG with RSOI. The electrons are also subject to a 1D periodic potential along the wire direction and
the in-plane magnetic field perpendicular to the wire. An orbital coupling between this originally isolated system and an
external environment causes dissipative processes affecting indirectly the spin dynamics through RSOI.

An external ac driving originates in our work from an applied ac electric field. We show that for such a driving the net
stationary charge current is strongly suppressed if the transport is governed only by electrons of the Bloch sub-bands
related to the same Bloch band which would result from the corresponding truly 1D problem without RSOI. However, at the
same time and under the same conditions a net stationary spin current turns out to be activated in a spatially asymmetric
situation and for finite values of the spin-orbit coupling strength and the coupling strength between the orbital degrees
of freedom. The magnetic field does not destroy this picture, but it can partly reduce or on the contrary enhance the
ratchet effect.

The paper is organized as follows. In Section \ref{formulation} we describe a model within which a ratchet like behavior
of the spin transport can be achieved and present a master equation in terms of populations and transition rates between
the basis states used to calculate the charge and spin currents. These basis states are then thoroughly discussed in
Section \ref{Diag_sigma_x}. A tight-binding model is formulated in Section \ref{sigma_DVR_TB_model}. In Section
\ref{transition_rates} we present the transition rates and their properties. Finally, in Section \ref{currents} we derive
{\it analytical} expressions for the charge and spin currents and explore the spin ratchet effect in the system.

\section{Formulation of the problem}\label{formulation}
The full Hamiltonian of our problem is
\begin{equation}
\hat{H}_{\mathrm{full}}(t)=\hat{H}+\hat{H}_{\mathrm{ext}}(t)+\hat{H}_\mathrm{bath},
\label{full_hamiltonian}
\end{equation}
where $\hat{H}$ is the Hamiltonian of the isolated periodic system, $\hat{H}_{\mathrm{ext}}(t)$ describes an external
driving and $\hat{H}_\mathrm{bath}$ represents the term responsible for dissipative processes.

The isolated quasi-1D periodic system is formed in a 2DEG ($x-z$ plane) with RSOI using a periodic potential along the
$x$-axis and a harmonic confinement along the $z$-axis. The whole system is in a uniform stationary magnetic field along
the $z$-axis:
\begin{equation}
\begin{split}
\hat{H}=\frac{\hbar^2\hat{\vec{k}}^2}{2m}&+\frac{m\omega_0^2\hat{z}^2}{2}-
\frac{\hbar^2k_{\mathrm{so}}}{m}\bigl(\hat{\sigma}_x\hat{k}_z-\hat{\sigma}_z\hat{k}_x\bigl)+\\
&+U(\hat{x})\biggl(1+\gamma\frac{\hat{z}^2}{L^2}\biggl)-g\mu_\mathrm{B}\hat{\sigma}_zH_0,
\end{split}
\label{isolated_hamiltonian}
\end{equation}
where $H_0$ is the $z$-component of the magnetic field $\vec{H}_0=(0,0,H_0)$, and we have used the gauge in which the
components of the vector potential are $A_x=-H_0y$, $A_y=A_z=0$ (Landau gauge). Additionally, we have taken into account
the fact that in a 2DEG $y=0$. In Eq. (\ref{isolated_hamiltonian}) the operator $\hat{\vec{k}}$ is related to the momentum
operator $\hat{\vec{p}}$ as $\hat{\vec{p}}=\hbar\hat{\vec{k}}$, $\omega_0$ is the harmonic confinement strength,
$k_\mathrm{so}$ the spin-orbit interaction strength, $\gamma$ the strength of the coupling between the orbital degrees of
freedom $x$ and $z$, $g$ the electron spin $g$-factor, $\mu_\mathrm{B}$ the Bohr magneton, and $U(\hat{x})$ denotes the
periodic potential with period $L$,
\begin{equation}
U(x+L)=U(x).
\label{periodic_potential}
\end{equation}

In the following we assume that the periodic structure is subject to an external homogeneous time-dependent electric
field. Only the $x$-component of the electric field vector is non-zero, that is the electric field is
parallel or anti-parallel to the $x$-axis. Experimentally this can be implemented using for example linearly polarized
light. The external force thus couples only to the $x$-component of the electron coordinate operator:
\begin{equation}
\hat{H}_{\mathrm{ext}}=-F(t)\hat{x},
\label{driving_hamiltonian}
\end{equation}
where the force $F(t)$ is unbiased. In this work we use the time-dependence
\begin{equation}
F(t)=F\cos(\Omega(t-t_0)).
\label{driving_force}
\end{equation}
The term "unbiased external force" should not be confused with voltage bias. An external force is called unbiased if it is
periodic in time and its mean value, that is its average over one period, is equal to zero. It is obviously our case as
one can see from Eq. (\ref{driving_force}).

The system is also coupled to an external bath. In the present work we assume the transverse confinement to be strong
enough so that the probabilities of direct bath-excited transitions between the transverse modes are negligibly small. In
other words, the wire is truly 1D from the point of view of the bath which directly changes only the dynamics along the
wire. Thus in our model the external environment couples to the electronic degrees of freedom only through $\hat{x}$. The
bath itself as well as its interaction with the quantum wire are described within the Caldeira-Leggett model
\cite{Caldeira,Caldeira_1},
\begin{equation}
\hat{H}_\mathrm{bath}=\frac{1}{2}\sum_{\alpha=1}^{N_O}\biggl[\frac{\hat{p}_\alpha^2}{m_\alpha}+
m_\alpha\omega_\alpha^2\biggl(\hat{x}_\alpha-\frac{c_\alpha}{m_\alpha\omega_\alpha^2}\hat{x}\biggl)^2\biggl].
\label{bath_hamiltonian}
\end{equation}
The bath is fully characterized by its spectral density defined as
\begin{equation}
J(\omega)\equiv\frac{\pi}{2}\sum_{\alpha=1}^{N_O}\frac{c_\alpha^2}{m_\alpha\omega_\alpha}\delta(\omega-\omega_\alpha).
\label{spectral_density}
\end{equation}
It is important to emphasize that, due to the spin-orbit interaction and orbit-orbit coupling, the direct dissipative
interaction between the longitudinal dynamics in the wire and the external environment has an indirect impact on the
transition rates between different transverse modes. The transverse dynamics in the wire indirectly feels the presence of
the external bath through the spin-orbit interaction and orbit-orbit coupling.

The dynamical quantities of interest are the charge and spin currents. Specifically, the longitudinal charge current
$J_\mathrm{C}(t)$ is given (see for example Ref.~\onlinecite{Grifoni}) as a statistical average of the longitudinal charge
current operator $\hat{J}_\mathrm{C}(t)$, {\it i.e.} the product of the velocity operator $\hat{v}(t)$ and the elementary
electronic charge $-e$,
\begin{equation}
\hat{J}_\mathrm{C}(t)=-e\hat{v}(t),
\label{charge_current_op}
\end{equation}
\begin{equation}
J_\mathrm{C}(t)=-e\frac{d}{dt}\mathrm{Tr}[\hat{x}\hat{\rho}(t)],
\label{charge_current}
\end{equation}
where $\hat{\rho}(t)=\mathrm{Tr}_\mathrm{bath}\hat{W}(t)$ is the reduced statistical operator of the system, that is the
full one $\hat{W}(t)$ with the bath degrees of freedom traced out.

For the longitudinal spin current operator we use the definition suggested by Shi {\it et al.}\cite{Shi},
\begin{equation}
\hat{J}_\mathrm{S}^i(t)=\frac{d}{dt}\bigl(\hat{\sigma}_i\hat{x}\bigl),
\label{spin_current_op}
\end{equation}
which was further developed and applied to a two-dimensional hole gas by Zhang {\it et al.}\cite{Zhang} The advantage of
this definition over the conventional one ($\hat{J}_\mathrm{S}^i=\hat{\sigma}_i\hat{v}$) is that using the corresponding
spin current,
\begin{equation}
J_\mathrm{S}^i(t)=\frac{d}{dt}\mathrm{Tr}\bigl(\hat{\sigma}_i\hat{x}\hat{\rho}(t)\bigl),
\label{spin_current}
\end{equation}
the continuity equation for the spin density can often be written without a source term, which means that the spin current
defined in this way is conserved. This conserved spin current can be uniquely related to the spin accumulation at a sample
boundary. The out-of-plane polarized spin accumulation can experimentally be measured with Kerr rotation microscopy
\cite{Sih} or the Faraday rotation technique \cite{Kato_1}. The in-plane spin polarization is not directly measured by
Kerr rotation microscopy, but it can still be scanned by a magneto-optic Kerr microscope using, {\it e.g.}, the cleaved
edge technology as discussed by Kotissek {\it et al.}\cite{Kotissek} Even when the continuity equation contains a source
term, there is still one advantage of the spin current operator definition (\ref{spin_current_op}). This definition leads
to a very reasonable physical result: the corresponding spin current in (\ref{spin_current}) vanishes in insulators. In
Section \ref{currents} we will return to this point and analytically prove that when the periodic potential gets stronger
and as a result the energy bands get narrower, that is when the system turns into an insulator, the spin current given by
Eq. (\ref{spin_current}) goes to zero. Below we will calculate only the spin current polarized along the $z$-axis and
denote this current as $J_\mathrm{S}$, {\it i.e.}, $J_\mathrm{S}(t)\equiv J_\mathrm{S}^z(t)$. The components of the spin
current polarized along the $x$ and $y$ axes are zero as shown in Appendix \ref{spin_current_x_y_conv_SC}. The discussion
of the difference between the conventional definition of the spin current and the spin current definition used in our work
can also be found in Appendix \ref{spin_current_x_y_conv_SC}.

It is convenient to calculate the traces in (\ref{charge_current}) and (\ref{spin_current}) using the basis which
diagonalizes both $\hat{x}$ and $\hat{\sigma}_z$, because this requires to determine only the diagonal elements of the
reduced density matrix. In a quasi-1D periodic system with RSOI the energy spectrum can be related to the one of the
corresponding truly 1D problem without RSOI\cite{Smirnov_1}. This links the Bloch bands of that truly 1D problem to the
Bloch sub-bands of the quasi-1D problem. The general structure of the results obtained in Ref.~\onlinecite{Smirnov_1} is
retained in the presence of the orbit-orbit coupling and a uniform stationary magnetic field along the $z$-axis. A slight
change of the theory is given in Appendix \ref{energy_spinors_magnetic_field}. We shall consider electron transport under
such conditions when only a finite number of the Bloch sub-bands is involved. The basis which diagonalizes $\hat{x}$ and
$\hat{\sigma}_z$ becomes in this case discrete. The total number of the Bloch sub-bands is equal to the product of the
number, $N_\mathrm{B}$, of the Bloch bands from the corresponding truly 1D problem without magnetic field and without
spin-orbit coupling, the number, $N_\mathrm{t}$, of the transverse modes and the number of the spin states. In this work we
shall use the model with $N_\mathrm{B}=1$, $N_\mathrm{t}=2$. Since there are only two spin states, the total number of the
Bloch sub-bands in our problem is equal to four. The representation in terms of the eigen-states of the coordinate
operator for a model with discrete $x$-values is called discrete variable representation (DVR) \cite{Harris}. Let us call
$\sigma$-DVR the representation in which both the coordinate and spin operators are diagonal. Denoting the $\sigma$-DVR
basis states as $\{|\alpha\rangle\}$ and eigen-values of $\hat{x}$ and $\hat{\sigma}_z$ in a state $|\alpha\rangle$ through
$x_\alpha$ and $\sigma_\alpha$, respectively, the charge and spin currents (\ref{charge_current}) and (\ref{spin_current})
are rewritten as
\begin{equation}
\begin{split}
&J_\mathrm{C}(t)=-e\sum_\alpha x_\alpha\frac{d}{dt}P_\alpha(t),\\
&J_\mathrm{S}(t)=\sum_\alpha\sigma_\alpha x_\alpha\frac{d}{dt}P_\alpha(t),
\end{split}
\label{charge_spin_currents_sigma_DVR}
\end{equation}
where $P_\alpha(t)\equiv\langle\alpha|\hat{\rho}(t)|\alpha\rangle$ is the population of the $\sigma$-DVR state
$|\alpha\rangle$ at time $t$.

We are interested in the long-time limit of the currents $\bar{J}_\mathrm{C}(t)$ and $\bar{J}_\mathrm{S}(t)$ averaged over
the driving period $T=2\pi/\Omega$ with the time average of a time dependent function $f(t)$ defined as
$\bar{f}(t)\equiv(1/T)\int_t^{t+T}dt'f(t')$. From (\ref{charge_spin_currents_sigma_DVR}) it follows
\begin{equation}
\begin{split}
&\bar{J}_\mathrm{C}(t)=-e\sum_\alpha x_\alpha\frac{d}{dt}\bar{P}_\alpha(t),\\
&\bar{J}_\mathrm{S}(t)=\sum_\alpha\sigma_\alpha x_\alpha\frac{d}{dt}\bar{P}_\alpha(t).
\end{split}
\label{averaged_currents}
\end{equation}

The advantage of working in the $\sigma$-DVR basis is that real-time path integral techniques can be used to exactly trace
out the bath degrees of freedom \cite{Grifoni_1,Weiss}. Moreover, at driving frequencies larger than the ones
characterizing the internal dynamics of the quasi-1D system coupled to the bath, the averaged populations
$\bar{P}_\alpha(t)$ can be found from the master equation,
\begin{equation}
\frac{d}{dt}\bar{P}_\alpha(t)=\sum_{\substack{\beta\\(\beta\neq\alpha)}}\bar{\Gamma}_{\alpha\beta}\bar{P}_\beta(t)-
\sum_{\substack{\beta\\(\beta\neq\alpha)}}\bar{\Gamma}_{\beta\alpha}\bar{P}_\alpha(t),
\label{averaged_master_equation}
\end{equation}
valid at long times. In Eq. (\ref{averaged_master_equation}) $\bar{\Gamma}_{\alpha\beta}$ is an averaged transition rate
from the state $|\beta\rangle$ to the state $|\alpha\rangle$. In order to obtain concrete expressions for the averaged
currents the $\sigma$-DVR basis must be found explicitly. This is the subject of the next section.

\section{Diagonalization of $\hat{\sigma}_z$ and $\hat{x}$: the $\sigma$-DVR basis}\label{Diag_sigma_x}
The eigen-states of the $\hat{\sigma}_z$ operator were found in Ref.~\onlinecite{Smirnov_1} (see Eq. (12) therein) for a
model without coupling between the orbital degrees of freedom and magnetic field. The changes necessary to include those
two effects are discussed in Appendix \ref{energy_spinors_magnetic_field}. The eigen-value equation for the
$\hat{\sigma}_z$ operator is
\begin{equation}
\hat{\sigma}_z|l,k_\mathrm{B},j,\sigma\rangle_{\gamma,j}=\sigma|l,k_\mathrm{B},j,\sigma\rangle_{\gamma,j}.
\label{eigen_value_eq_sigma}
\end{equation}
In Eq. (\ref{eigen_value_eq_sigma}) $l$, $k_\mathrm{B}$, $j$, $\sigma$ stand for the Bloch band index, quasi-momentum,
transverse mode index and $z$-projection of the spin, respectively. Since in the presence of the orbit-orbit coupling the
periodic potential $U_{\gamma,j}(x)$ (see Appendix \ref{energy_spinors_magnetic_field}) depends on $\gamma$ and $j$, we
have labeled the ket-symbol with the subscript $\gamma,j$. In the ensuing analysis we follow the same rule and label all
the bra- and ket-symbols with the subscript $\gamma,j$, that is $_{\gamma,j}\langle\cdots|$ and $|\cdots\rangle_{\gamma,j}$.

It is convenient to start the diagonalization of the coordinate operator writing its matrix in the
$\{|l,k_\mathrm{B},j,\sigma\rangle_{\gamma,j}\}$ representation:
\begin{equation}
\begin{split}
&_{\gamma,j'}\langle l',k_\mathrm{B}',j',\sigma'|\hat{x}|l,k_\mathrm{B},j,\sigma\rangle_{\gamma,j}=\\
&=\delta_{j',j}\delta_{\sigma',\sigma}\;\;
_{\gamma,j}\langle l',k_\mathrm{B}'+\sigma k_\mathrm{so}|\hat{x}|l,k_\mathrm{B}+\sigma k_\mathrm{so}\rangle_{\gamma,j}.
\end{split}
\label{x_l_kb_j_sigma}
\end{equation}
The diagonal blocks,
\begin{equation}
\begin{split}
&_{\gamma,j}\langle l',k_\mathrm{B}',j,\sigma=1|\hat{x}|l,k_\mathrm{B},j,\sigma=1\rangle_{\gamma,j}=\\
&=\,_{\gamma,j}\langle l',k_\mathrm{B}'+k_\mathrm{so}|\hat{x}|l,k_\mathrm{B}+k_\mathrm{so}\rangle_{\gamma,j},\quad
\forall\,j,\\
&_{\gamma,j}\langle l',k_\mathrm{B}',j,\sigma=-1|\hat{x}|l,k_\mathrm{B},j,\sigma=-1\rangle_{\gamma,j}=\\
&=\,_{\gamma,j}\langle l',k_\mathrm{B}'-k_\mathrm{so}|\hat{x}|l,k_\mathrm{B}-k_\mathrm{so}\rangle_{\gamma,j},\quad \forall\,j,
\end{split}
\label{x_l_kb_j_sigma_diag_blocks}
\end{equation}
are unitary equivalent for a given value of the index $j$ and thus the eigen-values of $\hat{x}$ do not depend on
$\sigma$.

As it is shown in Appendix \ref{eigen_value_x}, the eigen-values of the matrix
$_{\gamma,j}\langle l',k_\mathrm{B}'|\hat{x}|l,k_\mathrm{B}\rangle_{\gamma,j}$ are
\begin{equation}
x_{\gamma;\zeta,m,j}=mL+d_{\gamma;\zeta,j},
\label{x_eigen_values}
\end{equation}
where $m=0,\pm1,\pm2\ldots$, $\zeta=1,2,\ldots,N_\mathrm{B}$ and the eigen-values $d_{\gamma;\zeta,j}$ are distributed within
one elementary cell. If, for example, the system is divided into the elementary cells in such a way that the origin of
coordinates is at the center of an elementary cell, then $-L/2<d_{\gamma;\zeta,j}\leqslant L/2$. In Eq.
(\ref{x_eigen_values}) we have taken into account that the periodic potential $U_{\gamma,j}(x)$, introduced in Appendix
\ref{energy_spinors_magnetic_field}, depends on $\gamma$ and $j$, and thus the eigen-values distributed within one
elementary cell also acquire a dependence on $\gamma$ and $j$.

From (\ref{x_l_kb_j_sigma}) and (\ref{x_eigen_values}) it follows that one can label the eigen-states of $\hat{x}$ with
the quantum numbers $\zeta$, $m$, $j$, $\sigma$, that is as $|\zeta,m,j,\sigma\rangle_{\gamma,j}$, and in the
$\{|l,k_\mathrm{B},j,\sigma\rangle_{\gamma,j}\}$ representation these eigen-states have the form:
\begin{equation}
\begin{split}
&_{\gamma,j'}\langle l,k_\mathrm{B},j',\sigma'|\zeta,m,j,\sigma\rangle_{\gamma,j}=\\
&=\delta_{j',j}\delta_{\sigma',\sigma}\;\;_{\gamma,j}\langle l,k_\mathrm{B},j,\sigma|\zeta,m,j,\sigma\rangle_{\gamma,j}.
\end{split}
\label{x_eigen_states}
\end{equation}
The corresponding eigen-values are $x_{\gamma;\zeta,m,j,\sigma}=x_{\gamma;\zeta,m,j}$. From the eigen-value equation
\begin{equation}
\hat{x}|\zeta,m,j,\sigma\rangle_{\gamma,j}=x_{\gamma;\zeta,m,j}|\zeta,m,j,\sigma\rangle_{\gamma,j}
\label{eigen_value_eq_x_op}
\end{equation}
written in the $\{|l,k_\mathrm{B},j,\sigma\rangle_{\gamma,j}\}$ representation through the use of (\ref{x_l_kb_j_sigma}),
\begin{equation}
\begin{split}
&\sum_{l',k_\mathrm{B}'}\,
_{\gamma,j}\langle l,k_\mathrm{B}+\sigma k_\mathrm{so}|\hat{x}|l',k_\mathrm{B}'+\sigma k_\mathrm{so}\rangle_{\gamma,j}\times\\
&\times\,_{\gamma,j}\langle l',k_\mathrm{B}',j,\sigma|\zeta,m,j,\sigma\rangle_{\gamma,j}=\\
&=x_{\gamma;\zeta,m,j}\;\;_{\gamma,j}\langle l,k_\mathrm{B},j,\sigma|\zeta,m,j,\sigma\rangle_{\gamma,j},
\end{split}
\label{eigen_value_eq_x_l_kb_j_sigma}
\end{equation}
it follows that
\begin{equation}
\begin{split}
&_{\gamma,j}\langle l,k_\mathrm{B},j,\sigma=1|\zeta,m,j,\sigma=1\rangle_{\gamma,j}=\\
&=\,_{\gamma,j}\langle l,k_\mathrm{B}+k_\mathrm{so}|\zeta,m\rangle_{\gamma,j},\\
&_{\gamma,j}\langle l,k_\mathrm{B},j,\sigma=-1|\zeta,m,j,\sigma=-1\rangle_{\gamma,j}=\\
&=\,_{\gamma,j}\langle l,k_\mathrm{B}-k_\mathrm{so}|\zeta,m\rangle_{\gamma,j}.
\end{split}
\label{x_eigen_states_1}
\end{equation}

Since $|\zeta,m,j,\sigma\rangle_{\gamma,j}$ is also the eigen-state of $\hat{\sigma}_z$ corresponding to the eigen-value
$\sigma_{\zeta,m,j,\sigma}=\sigma$, we infer that the $\sigma$-DVR basis states $|\alpha\rangle$ from the previous section
are just the $|\zeta,m,j,\sigma\rangle_{\gamma,j}$ states, that is
$\{|\alpha\rangle\}\equiv\{|\zeta,m,j,\sigma\rangle_{\gamma,j}\}$.

\section{$\sigma$-DVR representation and its tight-binding model}\label{sigma_DVR_TB_model}
Let us represent the Hamiltonian $\hat{H}$ in the $\sigma$-DVR basis obtained in the previous section in order to derive
an effective tight-binding model.

Using the $\{|\zeta,m,j,\sigma\rangle_{\gamma,j}\}$ basis the Hamiltonian $\hat{H}$ can be written as
\begin{equation}
\begin{split}
&\hat{H}=
\sum_{\substack{\zeta,m,j,\sigma\\\zeta',m',j',\sigma'}}\,
_{\gamma,j'}\langle\zeta',m',j',\sigma'|\hat{H}|\zeta,m,j,\sigma\rangle_{\gamma,j}\times\\
&\times|\zeta',m',j',\sigma'\rangle_{\gamma,j'}\;\;_{\gamma,j}\langle\zeta,m,j,\sigma|,
\end{split}
\label{H_sigma_DVR}
\end{equation}
with the matrix
\begin{equation}
\begin{split}
&_{\gamma,j'}\langle\zeta',m',j',\sigma'|\hat{H}|\zeta,m,j,\sigma\rangle_{\gamma,j}=\sum_{l,k_\mathrm{B},\eta}
\varepsilon_{\gamma;l,\eta}(k_\mathrm{B})\times\\
&\times\,_{\gamma,j'}\langle\zeta',m'|l,k_\mathrm{B}+\sigma'k_\mathrm{so}\rangle_{\gamma,j'}\times\\
&\times\,_{\gamma,j}\langle l,k_\mathrm{B}+\sigma k_\mathrm{so}|\zeta,m\rangle_{\gamma,j}\;
\theta_{\gamma;l,k_\mathrm{B},\eta}(j',\sigma')\times\\
&\times\theta_{\gamma;l,k_\mathrm{B},\eta}^*(j,\sigma).
\end{split}
\label{matrix_H_sigma_DVR}
\end{equation}

The tight-binding approximation of (\ref{H_sigma_DVR}) is obtained if one assumes that the matrix elements
(\ref{matrix_H_sigma_DVR}) with $|m'-m|>1$ are negligibly small.

We consider temperatures low enough and assume that electrons populate only the lowest Bloch sub-bands with $l=1$
({\it i.e.,} $N_\mathrm{B}=1$). Under this condition the periodic potential can be of arbitrary shape and the only
limitation on it is the validity of the tight-binding approximation.

Below we thoroughly study the case where the four lowest Bloch sub-bands are the ones with $l=1,\,\eta=1,2,3,4$ and the
only ones which are populated with electrons. For simplicity we consider weak orbit-orbit coupling and calculate the
corresponding eigen-energies $\varepsilon_{\gamma;l,\eta}(k_\mathrm{B})$ and eigen-spinors
$\theta_{\gamma;l,k_\mathrm{B},\eta}(j,\sigma)$ retaining only the first two transverse modes, that is $j=0,1$. In this case
$\hat{H}$ has the form
\begin{equation}
\begin{split}
&\hat{H}=\sum_m\biggl[\sum_{j,\sigma}\varepsilon_{\gamma;j,\sigma}|m,j,\sigma\rangle_{\gamma,j}\;
_{\gamma,j}\langle m,j,\sigma|+\\
&+\!\!\!\!\sum_{j,\sigma'\neq\sigma}\!\!\!\Delta^\mathrm{intra}_{\gamma;j,\sigma';j,\sigma}(m)|m,j,\sigma'\rangle_{\gamma,j}\;
_{\gamma,j}\langle m,j,\sigma|+\\
&+\!\!\!\!\!\sum_{j'\neq j,\sigma',\sigma}\!\!\!\!\!\!
\Delta^\mathrm{intra}_{\gamma;j',\sigma';j,\sigma}(m)|m,j',\sigma'\rangle_{\gamma,j'}\;_{\gamma,j}\langle m,j,\sigma|+\\
&+\!\!\!\!\!\!\!\sum_{j',j,\sigma',\sigma}\!\!\!\biggl(\!\!
\Delta^\mathrm{inter,b}_{\gamma;j',\sigma';j,\sigma}(m)|m,j',\sigma'\rangle_{\gamma,j'}\;_{\gamma,j}\langle m+1,j,\sigma|+\\
&+\Delta^\mathrm{inter,f}_{\gamma;j',\sigma';j,\sigma}(m)|m+1,j',\sigma'\rangle_{\gamma,j'}\;
_{\gamma,j}\langle m,j,\sigma|\biggl)\biggl],
\end{split}
\label{H_tb_one_band}
\end{equation}
where
\begin{equation}
|m,j,\sigma\rangle_{\gamma,j}\equiv|\zeta=1,m,j,\sigma\rangle_{\gamma,j},
\label{x_states_short}
\end{equation}
and we have defined the on-site energies $\varepsilon_{\gamma;j,\sigma}$ and hopping matrix elements
$\Delta^\mathrm{intra}_{\gamma;j',\sigma';j,\sigma}(m)$, $\Delta^\mathrm{inter,b}_{\gamma;j',\sigma';j,\sigma }(m)$ and
$\Delta^\mathrm{inter,f}_{\gamma;j',\sigma';j,\sigma}(m)$ as follows
\begin{equation}
\begin{split}
&\varepsilon_{\gamma;j,\sigma}\equiv\,_{\gamma,j}\langle m,j,\sigma|\hat{H}|m,j,\sigma\rangle_{\gamma,j},\\
&\Delta^\mathrm{intra}_{\gamma;j',\sigma';j,\sigma}(m)\!\!\!\underset{(j',\sigma')\neq(j,\sigma)}{\equiv}\!\!\!\,
_{\gamma,j'}\langle m,j',\sigma'|\hat{H}|m,j,\sigma\rangle_{\gamma,j},\\
&\Delta^\mathrm{inter,b}_{\gamma;j',\sigma';j,\sigma}(m)\equiv\,
_{\gamma,j'}\langle m,j',\sigma'|\hat{H}|m+1,j,\sigma\rangle_{\gamma,j},\\
&\Delta^\mathrm{inter,f}_{\gamma;j',\sigma';j,\sigma}(m)\equiv\,
_{\gamma,j'}\langle m+1,j',\sigma'|\hat{H}|m,j,\sigma\rangle_{\gamma,j}.
\end{split}
\label{H_tb_param}
\end{equation}
Note that
\begin{equation}
[\Delta^\mathrm{intra}_{\gamma;j',\sigma';j,\sigma}(m)]^*=\Delta^\mathrm{intra}_{\gamma;j,\sigma;j',\sigma'}(m),
\label{relations_hop_intra}
\end{equation}
\begin{equation}
[\Delta^\mathrm{inter,b}_{\gamma;j',\sigma';j,\sigma}(m)]^*=\Delta^\mathrm{inter,f}_{\gamma;j,\sigma;j',\sigma'}(m).
\label{relations_hop_inter}
\end{equation}
Introducing the notations
\begin{equation}
\begin{split}
&\{\xi\}\equiv\{(j,\sigma)\},\\
&\xi=1\Leftrightarrow(0,1),\,\xi=2\Leftrightarrow(0,-1),\\
&\xi=3\Leftrightarrow(1,1),\,\xi=4\Leftrightarrow(1,-1),
\end{split}
\label{j_sigma_to_xi}
\end{equation}
we finally have
\begin{equation}
\begin{split}
&\hat{H}=\sum_m\biggl[\sum_{\xi=1}^4\varepsilon_{\gamma;\xi}|m,\xi\rangle_{\gamma,\xi}\;_{\gamma,\xi}\langle m,\xi|+\\
&+\sum_{\xi\neq\xi'=1}^4\Delta^\mathrm{intra}_{\gamma;\xi',\xi}(m)|m,\xi'\rangle_{\gamma,\xi'}\;_{\gamma,\xi}\langle m,\xi|+\\
&+\sum_{\xi,\xi'=1}^4\biggl(\Delta^\mathrm{inter,b}_{\gamma;\xi',\xi}(m)|m,\xi'\rangle_{\gamma,\xi'}\;
_{\gamma,\xi}\langle m+1,\xi|+\\
&+\Delta^\mathrm{inter,f}_{\gamma;\xi',\xi}(m)|m+1,\xi'\rangle_{\gamma,\xi'}\;_{\gamma,\xi}\langle m,\xi|\biggl)\biggl].
\end{split}
\label{H_tb_final}
\end{equation}
Equation (\ref{H_tb_final}) represents a tight-binding model which can now be used to perform actual calculations of
quantum transport in a dissipative system.

To conclude this section, we would like to note that because of the simultaneous presence of the harmonic confinement and
RSOI the system splits into two subsystems. The first subsystem is characterized by $\xi=1,4$ and the second one by
$\xi=2,3$. These subsystems are totally decoupled: there is no electron exchange between them. Such a state of affairs
persists if one considers more than two transverse modes. In this work, for simplicity, we only consider one subsystem,
namely the one with $\xi=1,4$. Such uncoupled subsystems also appear within the hard wall model of the transverse
confinement \cite{Perroni}.

\section{Transition rates}\label{transition_rates}
The tight-binding model introduced in Section \ref{sigma_DVR_TB_model} relies upon the fact that the hopping matrix
elements (\ref{H_tb_param}) are small. In this case the second-order approximation for the averaged transition rates in
Eq. (\ref{averaged_master_equation}) can be used giving \cite{Grifoni,Hartmann}
\begin{equation}
\begin{split}
&\bar{\Gamma}_{\gamma;\xi',\xi}^{m',m}=\frac{|\Delta_{\gamma;\xi',\xi}^{m',m}|^2}{\hbar^2}\times\\
\times\int_{-\infty}^{\infty}\!\!\!\!\!\!d\tau&\mathrm{e}^{-[(x_{\gamma;m,\xi}-x_{\gamma;m',\xi'})^2/\hbar]
Q(\tau)+\mathrm{i}[(\varepsilon_{\gamma;\xi}-\varepsilon_{\gamma;\xi'})/\hbar]\tau}\times\\
&\times J_0\biggl[\frac{2F(x_{\gamma;m,\xi}-x_{\gamma;m',\xi'})}{\hbar\Omega}\sin\biggl(\frac{\Omega\tau}{2}\biggl)\biggl],
\end{split}
\label{transition_rate}
\end{equation}
where $x_{\gamma;m,\xi}\equiv x_{\gamma;\zeta=1,m,\xi}=mL+d_{\gamma;\xi}$ with $d_{\gamma;\xi}\equiv d_{\gamma;1,j}$,
$\Delta_{\gamma;\xi',\xi}^{m',m}\equiv\,_{\gamma,\xi'}\langle m',\xi'|\hat{H}|m,\xi\rangle_{\gamma,\xi}$ the hopping matrix
element between the states $|m',\xi'\rangle_{\gamma,\xi'}$ and $|m,\xi\rangle_{\gamma,\xi}$, $J_0(x)$ the zero-order Bessel
function and $Q(\tau)$ the twice integrated bath correlation function\cite{Weiss}:
\begin{equation}
\begin{split}
Q(\tau)=\frac{1}{\pi}\int_0^\infty\,&d\omega\frac{J(\omega)}{\omega^2}\biggl[\coth\biggl(\frac{\hbar\omega\beta}{2}\biggl)
\times\\
&\times[1-\cos(\omega\tau)]+\mathrm{i}\sin(\omega\tau)\biggl],
\end{split}
\label{tibcf}
\end{equation}
where $J(\omega)$ is given by Eq. (\ref{spectral_density}) and $\beta$ is the inverse temperature.

The transition rates are functions of the orbit-orbit coupling strength $\gamma$ because the Bloch amplitudes as well
as the difference $\Delta d_{\gamma}\equiv d_{\gamma;1,0}-d_{\gamma;1,1}$ depend on $\gamma$. Within the context of the
tight-binding model the eigen-values $d_{\gamma;1,0}$ and $d_{\gamma;1,1}$ tend to zero and fulfil
$\Delta d_{\gamma}/l_\mathrm{r}\ll 1$, where $l_\mathrm{r}=\mathrm{min}[L,\sqrt{\hbar/m\omega_0},\hbar\Omega/F,\ldots]$.
Consequently, the transition rates depend on $\gamma$ predominantly through the Bloch amplitudes, and in this work we pay
no regard to terms of order $\mathcal{O}(\Delta d_{\gamma}/l_\mathrm{r})$. This is also consistent with our model taking
into account only the first two transverse modes. Keeping terms of order $\mathcal{O}(\Delta d_{\gamma}/l_\mathrm{r})$ would
mean that the strength $\gamma$ of the orbit-orbit coupling is large enough so that one would need to consider more than
just the first two transverse modes because in this case the non-diagonal elements would be comparable with the diagonal
ones.

Using the notations,
\begin{equation}
\begin{split}
&\bar{\Gamma}_{\gamma;\xi',\xi}^{m,m}\equiv\bar{\Gamma}_{\gamma;\xi',\xi}^\mathrm{intra},\quad \xi'\neq\xi,\\
&\bar{\Gamma}_{\gamma;\xi',\xi}^{m,m+1}\equiv\bar{\Gamma}_{\gamma;\xi',\xi}^\mathrm{inter,b},\\
&\bar{\Gamma}_{\gamma;\xi',\xi}^{m+1,m}\equiv\bar{\Gamma}_{\gamma;\xi',\xi}^\mathrm{inter,f},
\end{split}
\label{intra_inter_notations}
\end{equation}
from (\ref{transition_rate}) one obtains
\begin{equation}
\bar{\Gamma}_{\gamma;\xi',\xi}^\mathrm{intra}=0,
\label{transition_rate_intra}
\end{equation}
and
\begin{equation}
\begin{split}
&\bar{\Gamma}_{\gamma;\xi',\xi}^\mathrm{inter,b}=|\Delta_{\gamma;\xi',\xi}^\mathrm{inter,b}(m)|^2J_{\gamma;\xi',\xi},\\
&\bar{\Gamma}_{\gamma;\xi',\xi}^\mathrm{inter,f}=|\Delta_{\gamma;\xi',\xi}^\mathrm{inter,f}(m)|^2J_{\gamma;\xi',\xi},
\end{split}
\label{transition_rate_inter}
\end{equation}
where
\begin{equation}
\begin{split}
&J_{\gamma;\xi',\xi}=\frac{1}{\hbar^2}
\int_{-\infty}^{\infty}d\tau\mathrm{e}^{-\frac{L^2}{\hbar}
Q(\tau)+\mathrm{i}[(\varepsilon_{\gamma;\xi}-\varepsilon_{\gamma;\xi'})/\hbar]\tau}\times\\
&\times J_0\biggl[\frac{2FL}{\hbar\Omega}\sin\biggl(\frac{\Omega\tau}{2}\biggl)\biggl].
\end{split}
\label{integral_J}
\end{equation}
Note that $\bar{\Gamma}_{\gamma;\xi',\xi}^\mathrm{inter,b}$ and $\bar{\Gamma}_{\gamma;\xi',\xi}^\mathrm{inter,f}$ do not depend
on $m$ due to the Bloch theorem which leads to an $m$-dependence of $\Delta_{\gamma;\xi',\xi}^\mathrm{inter,b}(m)$ and
$\Delta_{\gamma;\xi',\xi}^\mathrm{inter,f}(m)$ only through a phase factor as it is shown in Appendix
\ref{Bloch_states_and_DVR}. From (\ref{relations_hop_inter}) and (\ref{transition_rate_inter}) it follows that
\begin{equation}
\bar{\Gamma}_{\gamma;\xi,\xi}^\mathrm{inter,b}=\bar{\Gamma}_{\gamma;\xi,\xi}^\mathrm{inter,f},
\label{transition_rate_relations_a}
\end{equation}
\begin{equation}
\bar{\Gamma}_{\gamma;\xi',\xi}^\mathrm{inter,b}\bar{\Gamma}_{\gamma;\xi,\xi'}^\mathrm{inter,b}=
\bar{\Gamma}_{\gamma;\xi',\xi}^\mathrm{inter,f}\bar{\Gamma}_{\gamma;\xi,\xi'}^\mathrm{inter,f}.
\label{transition_rate_relations_b}
\end{equation}

To calculate the charge and spin currents we additionally need the transition rates
\begin{equation}
\bar{\Gamma}_{\gamma;\xi,\xi'}\equiv\bar{\Gamma}_{\gamma;\xi,\xi'}^\mathrm{inter,f}+
\bar{\Gamma}_{\gamma;\xi,\xi'}^\mathrm{intra}+\bar{\Gamma}_{\gamma;\xi,\xi'}^\mathrm{inter,b}.
\label{transition_rates_add}
\end{equation}
As pointed out at the end of Section \ref{sigma_DVR_TB_model}, the system is split into two subsystems isolated from each
other. Since electron exchange between the subsystems is absent one can write
\begin{equation}
\begin{split}
&\bar{\Gamma}_{\gamma;1,2}=\bar{\Gamma}_{\gamma;1,3}=\bar{\Gamma}_{\gamma;2,1}=\bar{\Gamma}_{\gamma;2,4}=\\
&=\bar{\Gamma}_{\gamma;3,1}=\bar{\Gamma}_{\gamma;3,4}=\bar{\Gamma}_{\gamma;4,2}=\bar{\Gamma}_{\gamma;4,3}=0.
\end{split}
\label{vanishing_transition_rates}
\end{equation}
The last equalities are very useful because they allow us to significantly simplify the expressions for the charge and
spin currents which are derived in the next section.
\\
\\

\section{Charge and spin currents}\label{currents}
The expressions for the stationary averaged charge and spin currents,
\begin{equation}
\bar{J}_\mathrm{C}^{\infty}\equiv\underset{t\to\infty}{\mathrm{lim}}\bar{J}_\mathrm{C}(t),\quad
\bar{J}_\mathrm{S}^{\infty}\equiv\underset{t\to\infty}{\mathrm{lim}}\bar{J}_\mathrm{S}(t),
\end{equation}
can be found from the averaged master equation (\ref{averaged_master_equation}) which we rewrite here using the
$\sigma$-DVR indices and tight-binding approximation introduced in Section \ref{sigma_DVR_TB_model} and utilizing the
notations of Section \ref{transition_rates} for the transition rates:
\begin{widetext}
\begin{equation}
\begin{split}
\frac{d}{dt}\bar{P}_{\gamma;\xi}^m(t)=
\!\!\sum_{\substack{\xi'=1\\(\xi'\neq\xi)}}^4&\bigl[\bar{\Gamma}_{\gamma;\xi,\xi'}^\mathrm{inter,f}\bar{P}_{\gamma;\xi'}^{m-1}(t)+
\bar{\Gamma}_{\gamma;\xi,\xi'}^\mathrm{intra}\bar{P}_{\gamma;\xi'}^m(t)+
\bar{\Gamma}_{\gamma;\xi,\xi'}^\mathrm{inter,b}\bar{P}_{\gamma;\xi'}^{m+1}(t)\bigl]-
\sum_{\substack{\xi'=1\\(\xi'\neq\xi)}}^4\bigl[\bar{\Gamma}_{\gamma;\xi',\xi}^\mathrm{inter,b}+
\bar{\Gamma}_{\gamma;\xi',\xi}^\mathrm{intra}+\bar{\Gamma}_{\gamma;\xi',\xi}^\mathrm{inter,f}\bigl]\bar{P}_{\gamma;\xi}^m(t)+\\
&+\bigl[\bar{\Gamma}_{\gamma;\xi,\xi}^\mathrm{inter,f}\bar{P}_{\gamma;\xi}^{m-1}(t)+
\bar{\Gamma}_{\gamma;\xi,\xi}^\mathrm{inter,b}\bar{P}_{\gamma;\xi}^{m+1}(t)\bigl]-
\bigl[\bar{\Gamma}_{\gamma;\xi,\xi}^\mathrm{inter,b}+\bar{\Gamma}_{\gamma;\xi,\xi}^\mathrm{inter,f}\bigl]
\bar{P}_{\gamma;\xi}^m(t),
\end{split}
\label{averaged_master_equation_full_sigma_DVR_tb}
\end{equation}
\end{widetext}
From (\ref{charge_spin_currents_sigma_DVR}) and (\ref{averaged_master_equation_full_sigma_DVR_tb}) one finds
\begin{equation}
\bar{J}_\mathrm{C}^\infty=-eL\sum_{\xi,\xi'=1}^4\bigl[\bar{\Gamma}_{\gamma;\xi,\xi'}^\mathrm{inter,f}-
\bar{\Gamma}_{\gamma;\xi,\xi'}^\mathrm{inter,b}\bigl]p_{\gamma;\xi'}^\infty,
\label{stationary_averaged_charge_current}
\end{equation}
\begin{equation}
\begin{split}
&\bar{J}_\mathrm{S}^\infty=\!\!\!\sum_{\xi,\xi'=1}^4\bigl[\bigl(d_{\gamma;\xi}\sigma_\xi-d_{\gamma;\xi'}\sigma_{\xi'}\bigl)
\bigl(\bar{\Gamma}_{\gamma;\xi,\xi'}^\mathrm{inter,f}+\bar{\Gamma}_{\gamma;\xi,\xi'}^\mathrm{inter,b}\bigl)+\\
&+L\sigma_\xi\bigl(\bar{\Gamma}_{\gamma;\xi,\xi'}^\mathrm{inter,f}-\bar{\Gamma}_{\gamma;\xi,\xi'}^\mathrm{inter,b}\bigl)\bigl]
p_{\gamma;\xi'}^\infty,
\end{split}
\label{stationary_averaged_spin_current}
\end{equation}
where we have used Eq. (\ref{x_eigen_values}). To derive Eq. (\ref{stationary_averaged_spin_current}) we have additionally
made use of Eq. (\ref{transition_rate_intra}). In Eq. (\ref{stationary_averaged_spin_current})
$\sigma_\xi\equiv\sigma_{\zeta=1,m,\xi}$ and $\sigma_1=\sigma_3=1,\,\,\sigma_2=\sigma_4=-1$ as it follows from
Eq. (\ref{j_sigma_to_xi}). The quantities $p_{\gamma;\xi}^\infty$ are defined as
\begin{equation}
p_{\gamma;\xi}(t)\equiv\sum_m\bar{P}_{\gamma;\xi}^m(t),\quad
p_{\gamma;\xi}^\infty\equiv\underset{t\to\infty}{\mathrm{lim}}p_{\gamma;\xi}(t),
\label{p_inf}
\end{equation}
and they satisfy the constraint
\begin{equation}
p_{\gamma;1}(t)+p_{\gamma;2}(t)+p_{\gamma;3}(t)+p_{\gamma;4}(t)=1,\quad\forall\,\,t.
\label{sum_of_all_probabilities}
\end{equation}
As already mentioned at the end of Section \ref{sigma_DVR_TB_model}, we only consider the subsystem with $\xi=1,4$. The
properties of the stationary averaged transport do not depend on initial conditions. We choose the following ones:
\begin{equation}
p_{\gamma;1}(t=0)=1,\quad p_{\gamma;4}(t=0)=0.
\label{initial_cond}
\end{equation}
Because of the constraint (\ref{sum_of_all_probabilities}) $p_{\gamma;2}(t=0)=p_{\gamma;3}(t=0)=0$ and since there is no
electron exchange between the subsystems, the states of the subsystem with $\xi=2,3$ remain empty at any time,
$p_{\gamma;2}(t)=p_{\gamma;3}(t)=0,\;\forall t$. This leads to $p_{\gamma;2}^\infty=p_{\gamma;3}^\infty=0$. Then from the master
equation (\ref{averaged_master_equation_full_sigma_DVR_tb}) with the initial conditions (\ref{initial_cond}) and using
(\ref{transition_rates_add}), (\ref{vanishing_transition_rates}) one obtains
\begin{equation}
p_{\gamma;1}^\infty=\frac{\bar{\Gamma}_{\gamma;1,4}}{\bar{\Gamma}_{\gamma;1,4}+\bar{\Gamma}_{\gamma;4,1}},\quad
p_{\gamma;4}^\infty=\frac{\bar{\Gamma}_{\gamma;4,1}}{\bar{\Gamma}_{\gamma;1,4}+\bar{\Gamma}_{\gamma;4,1}}.
\label{p_1_4}
\end{equation}
Using Eqs. (\ref{transition_rate_intra}), (\ref{transition_rate_relations_a})-(\ref{transition_rates_add}) and
(\ref{p_1_4}) it follows from (\ref{stationary_averaged_charge_current})
\begin{equation}
\bar{J}_\mathrm{C}^\infty=0,
\label{zero_stationary_averaged_charge_current}
\end{equation}
that is the absence of the stationary averaged charge transport. However, using Eqs. (\ref{transition_rate_intra}),
(\ref{transition_rate_relations_a}), (\ref{transition_rates_add}) and (\ref{p_1_4}) we get from Eq.
(\ref{stationary_averaged_spin_current})
\begin{equation}
\bar{J}_\mathrm{S}^\infty\!\!=\!\!
\frac{2L}{\bar{\Gamma}_{\gamma;1,4}+\bar{\Gamma}_{\gamma;4,1}}
\bigl(\bar{\Gamma}_{\gamma;1,4}^\mathrm{inter,f}\bar{\Gamma}_{\gamma;4,1}^\mathrm{inter,b}\!-
\bar{\Gamma}_{\gamma;1,4}^\mathrm{inter,b}\bar{\Gamma}_{\gamma;4,1}^\mathrm{inter,f}\bigl).
\label{stationary_averaged_spin_current_a}
\end{equation}
The last expression can be rewritten in terms of the hopping matrix elements $\Delta^\mathrm{inter,f}_{\gamma;\xi',\xi}(m)$.
Making use of Eqs. (\ref{relations_hop_inter}), (\ref{transition_rate_intra}), (\ref{transition_rate_inter}) and
(\ref{transition_rates_add}) we derive the stationary averaged spin current:
\begin{equation}
\begin{split}
&\bar{J}_\mathrm{S}^\infty=2L\frac{J_{\gamma;1,4}J_{\gamma;4,1}}{J_{\gamma;1,4}+J_{\gamma;4,1}}\times\\
&\times\bigl(|\Delta^\mathrm{inter,f}_{\gamma;1,4}(m)|^2-|\Delta^\mathrm{inter,f}_{\gamma;4,1}(m)|^2\bigl).
\end{split}
\label{stationary_averaged_spin_current_b}
\end{equation}
Using Eqs. (\ref{stationary_averaged_spin_current_b}) and (\ref{difference_of_deltas}) the non-equilibrium stationary
averaged spin current can be written as
\begin{equation}
\begin{split}
&\bar{J}_\mathrm{n-e,S}^\infty=-2\biggl(\frac{J_{\gamma;1,4}J_{\gamma;4,1}}{J_{\gamma;1,4}+J_{\gamma;4,1}}-
\frac{J^{(0)}_{\gamma;1,4}J^{(0)}_{\gamma;4,1}}{J^{(0)}_{\gamma;1,4}+J^{(0)}_{\gamma;4,1}}\biggl)\times\\
&\times\frac{L\hbar^3 k^2_\mathrm{so}\omega_0}{m}
\sum_{k_\mathrm{B},k_\mathrm{B}'}\sin[(k_\mathrm{B}-k_\mathrm{B}')L]\,\mathrm{Im}[F_{\gamma;k_\mathrm{B},k_\mathrm{B}'}],
\end{split}
\label{stationary_averaged_spin_current_c}
\end{equation}
where $J^{(0)}_{\gamma;\xi',\xi}$ is given by Eq. (\ref{integral_J}) with $F=0$ and the function
$F_{\gamma;k_\mathrm{B},k_\mathrm{B}'}$ is defined by Eq. (\ref{F_function}). Note the structure of Eq.
(\ref{stationary_averaged_spin_current_c}). It is the product of two factors of different physical origin. The factor in
the second line describes the isolated system and the factor in the first line is purely due to the dissipative coupling
to an external environment. To get Eq. (\ref{stationary_averaged_spin_current_c}) we have eliminated from
$\bar{J}_\mathrm{S}^\infty$ the equilibrium spin current arising due to the non-compensation\cite{Rashba_1} of the spin
currents from different bands of the Rashba-Bloch spectrum of the isolated system. It turns out that this effect is strong
enough to indenture in a dissipative system. Below we only consider the non-equilibrium spin current,
$\bar{J}_\mathrm{n-e,S}^\infty$, and not the full one, $\bar{J}_\mathrm{S}^\infty$.

Let us at this point also mention the dependence of the spin current $\bar{J}_\mathrm{n-e,S}^\infty$ on the magnetic field
$H_0$. Since the magnetic field is applied along the $z$-axis, it couples to the system through the $\hat{\sigma}_z$
operator and thus the hopping matrix elements $\Delta^\mathrm{inter,f}_{\gamma;1(4),4(1)}(m)$ do not depend on $H_0$. It then
follows that the spin current depends on the magnetic field only through its dissipative prefactor. The dependence on $H_0$
comes into play through the on-site energies $\varepsilon_{\gamma;1(4)}$. The difference
$\varepsilon_{\gamma;4}-\varepsilon_{\gamma;1}$ which enters the integrals $J_{\gamma;1(4),4(1)}$ and
$J^{(0)}_{\gamma;1(4),4(1))}$ can be written as:
\begin{equation}
\begin{split}
\varepsilon_{\gamma;4}-\varepsilon_{\gamma;1}=&\frac{1}{N}\sum_{k_\mathrm{B}}\bigl[\varepsilon_{\gamma,1;1}^{(0)}(k_\mathrm{B})-
\varepsilon_{\gamma,0;1}^{(0)}(k_\mathrm{B})\bigl]+\\
&+\hbar\omega_0+2g\mu_\mathrm{B}H_0,
\end{split}
\label{difference_on-site_energies}
\end{equation}
where $N$ is the number of the elementary cells and $\varepsilon_{\gamma,j;l}^{(0)}(k_\mathrm{B})$ are the eigen-values of
the truly 1D Hamiltonian
\begin{equation}
\hat{H}_{0;\gamma,j}^\mathrm{1D}\equiv\frac{\hbar^2\hat{k}_x^2}{2m}+
U(\hat{x})\biggl[1+\gamma\frac{\hbar}{m\omega_0L^2}\biggl(j+\frac{1}{2}\biggl)\biggl].
\label{hamiltonian_truly_1D}
\end{equation}

Therefore, in the presence of a transverse in-plane uniform stationary magnetic field the existence of the spin current is
possible under the same conditions which were discussed in Ref.~\onlinecite{Smirnov}. For completeness we list these
conditions below.

From (\ref{stationary_averaged_spin_current_c}) one finds, as mentioned in Section \ref{formulation}, that when the
electronic states become localized, the stationary averaged spin current vanishes. Indeed, in this insulating limit the
function $F_{\gamma;k_\mathrm{B},k_\mathrm{B}'}$ does not depend on the quasi-momenta $k_\mathrm{B}$ and $k_\mathrm{B}'$ and Eq.
(\ref{stationary_averaged_spin_current_c}) gives zero.

When the spin-orbit interaction is absent, that is $k_\mathrm{so}=0$, we get from
(\ref{stationary_averaged_spin_current_c})
\begin{equation}
\bar{J}_\mathrm{n-e,S}^\infty\bigl|_{k_\mathrm{so}=0}=0.
\label{spin_current_kso_0}
\end{equation}

Further, if the orbital degrees of freedom $x$ and $z$ are not coupled, that is $\gamma=0$, it follows from Eqs.
(\ref{stationary_averaged_spin_current_c}) and (\ref{Im_F_even}) that
\begin{equation}
\bar{J}_\mathrm{n-e,S}^\infty\bigl|_{\gamma=0}=0.
\label{spin_current_gamma_0}
\end{equation}

Finally, if the periodic potential is symmetric, the Bloch amplitudes are real and we find from Eqs.
(\ref{stationary_averaged_spin_current_c}) and (\ref{real_u_real_F})
\begin{equation}
\bar{J}_\mathrm{n-e,S}^\infty=0,\quad\mathrm{\it for\;symmetric\;periodic\;potentials}.
\label{spin_current_symmetric}
\end{equation}

Summarizing the results of this section we conclude that in order to generate a finite stationary averaged spin current
three conditions must simultaneously be fulfilled: 1) presence of the spin-orbit interaction in the isolated system;
2) finite coupling between the orbital degrees of freedom $x$ and $z$; 3) absence of the real space inversion center in
the isolated system.

Among these three conditions the second condition is perhaps less transparent and a simplified physical interpretation is
necessary. We propose the following physical explanation. The orbit-orbit coupling leads to the situation in which the
strength of the periodic potential varies across the quasi-1D wire. The periodic potential is equal to $U(x)$ in the
center of the wire and gets stronger closer to its edges. Thus the electron group velocity is larger in the central region
of the wire and decreases closer to the edges. At the same time the electron distribution across the channel depends on
the transverse mode $j$. It is given by the Hermite polynomials. For $j=0$ the electrons populate the center of the wire
while for $j=1$ they are distributed in regions closer to the edges. Hence, the electrons with $j=0$ are faster than those
with $j=1$. Because of the mixing between the confinement and RSOI different transverse modes carry different spin states.
Therefore, we conclude that different spin states have different group velocities along the wire. This difference results
in a finite longitudinal spin current.

Finally, one observes that a transverse in-plane uniform stationary magnetic field alone is not
enough to produce the spin current in a driven dissipative system. The magnetic field can only affect the magnitude of the
spin current when the properties of the isolated system meet the three conditions derived above.

\section{Results}\label{results}
In this section we show some results obtained numerically for the theoretical model developed in the previous sections. As
an example we consider an InGaAs/InP quantum wire structure. The values of the corresponding parameters used to get the
results are similar to the ones from the work of Sch\"apers {\it et al.}\cite{Schaepers} In particular,
$\hbar\omega_0=0.225$ meV, $\alpha\equiv \hbar^2 k_\mathrm{so}/m=9.94\times 10^{-12}$ eV$\cdot$m (which gives
$k_\mathrm{so}=4.82\times 10^6$ m$^{-1}$), $m=0.037m_0$ ($m_0$ is the free-electron mass). The value, $g=7.5$, of the
electron spin $g$-factor (in our notations $g\equiv -g^*/2$, where $g^*$ is the effective gyroscopic factor measured
experimentally) is taken from Ref.~\onlinecite{Madelung}. From these parameters and for example for the period of the
super-lattice $L=2.5\sqrt{\hbar/m\omega_0}\approx 0.24$ $\mu$m, which is easily achievable technologically at present
\cite{Steinshnider}, it follows that $k_\mathrm{so}L\approx 0.368\pi$.

\begin{figure}
\includegraphics[width=8.5 cm]{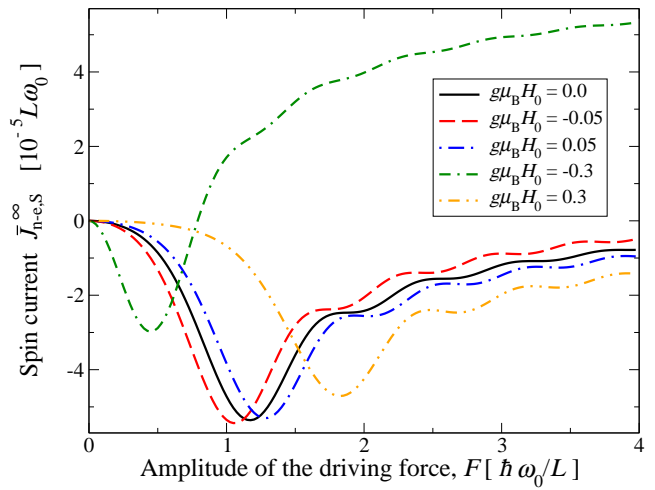}
\caption{\label{figure_1} (Color online) Non-equilibrium spin current, $\bar{J}^\infty_\mathrm{n-e,S}$, as a function of the
amplitude, $F$, of the driving force for different values of the $z$-projection of the magnetic field $H_0$. Further
parameters: temperature $k_\mathrm{Boltz.}T=0.5$, spin-orbit coupling strength $k_\mathrm{so}$ with $k_\text{so}L=\pi/2$,
orbit-orbit coupling strength $\gamma=0.08$, driving frequency $\Omega=0.2$, viscosity coefficient $\eta=0.08$.}
\end{figure}

The asymmetric periodic potential is
\begin{equation}
\begin{split}
U(x)=\hbar\omega_0\biggl\{&2.6\biggl[1-\cos\biggl(\frac{2\pi x}{L}-1.9\biggl)\biggl]+\\
&+1.9\cos\biggl(\frac{4\pi x}{L}\biggl)\biggl\}.
\end{split}
\label{asm_per_pot}
\end{equation}

The bath is assumed to be Ohmic with exponential cutoff:
\begin{equation}
J(\omega)=\eta\omega\exp\biggl(-\frac{\omega}{\omega_c}\biggl),
\label{Ohmic_spc_den}
\end{equation}
where $\eta$ is the viscosity coefficient and $\omega_c$ the cutoff frequency. We use $\omega_c=10\,\omega_0$.

To present the results we use in all the figures the units of $\hbar\omega_0$ and $\omega_0$ for energies and frequencies,
respectively. The viscosity coefficient is taken in units of $m\omega_0$.

\begin{figure}
\includegraphics[width=8.5 cm]{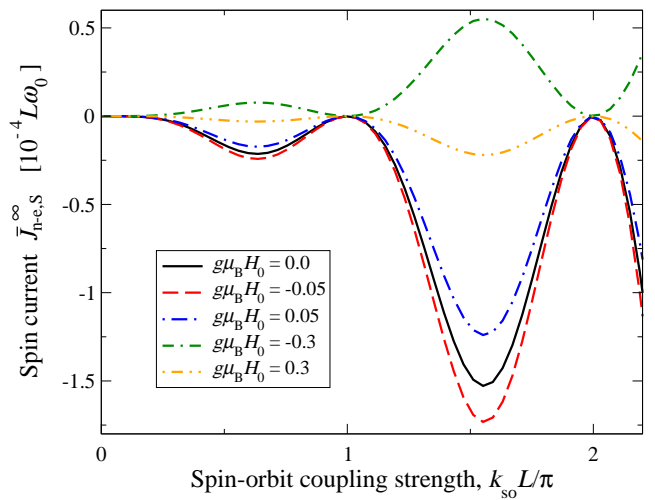}
\caption{\label{figure_2} (Color online) Non-equilibrium spin current, $\bar{J}^\infty_\mathrm{n-e,S}$, as a function of the
spin-orbit coupling strength, $k_\mathrm{so}$, for different values of the $z$-projection of the magnetic field $H_0$. The
driving amplitude is $F=1.0\,\hbar\omega_0/L$. The other parameters are as in Fig.~\ref{figure_1}.}
\end{figure}

Let us discuss possible values of the driving parameters. In a dissipationless system (or in a system with weak
dissipation) of size $L$ one should restrict possible values of the driving amplitude and frequency, $0<FL<\hbar\omega_0$
and $0<\Omega<\omega_0$, in order to stay within the validity of the model with the first two transverse modes opened. In a
strongly dissipative system, as in our case, it is not necessary to fulfil the last inequalities because an electron loses
a huge amount of its energy due to intensive dissipative processes. In general, our model of a driven strongly dissipative
system taking into account the first four Bloch sub-bands remains valid if at long times the electron energy averaged over
one period of the driving force, $\epsilon_{av}(F,\Omega,\eta)$ (which is a function of the driving and dissipation
parameters), is smaller than $\hbar\omega_0$, $\epsilon_{av}(F,\Omega,\eta)<\hbar\omega_0$. This can take place even if
$FL>\hbar\omega_0$ and $\Omega>\omega_0$ because even at such driving the strong dissipation (large values of $\eta$) will
consume major amount of the electron energy.

In Fig.~\ref{figure_1} the non-equilibrium spin current as a function of the amplitude of the external driving is shown
for different values of the $z$-projection of the magnetic field. For small values of the driving amplitude and small
magnetic fields it is seen that if the magnetic field has the same direction as the $z$-axis, the spin current decreases,
while the opposite direction of the magnetic field amplifies the spin current. This behavior can be physically understood
from Eq. (\ref{difference_on-site_energies}). Positive values of $H_0$ can be equivalently considered as larger values of
$\hbar\omega_0$, that is of the distance between the transverse modes. This in turn leads to a decrease of the transition
probabilities which suppresses the spin current. On the contrary, negative values of $H_0$ correspond to smaller values of
$\hbar\omega_0$ leading to an increase of the transition rates and thus the spin current is enhanced. Another physical
explanation is that the magnetic field aligns the spins along its direction. Therefore, when $H_0$ is positive or negative
the spins are forced to point in the direction of the $z$-axis or in the opposite direction, respectively. The spin
current gets more polarized in the direction of the $z$-axis for $H_0>0$ or in the opposite direction for $H_0<0$. As a
consequence its magnitude decreases for $H_0>0$ or increases for $H_0<0$ since it was polarized in the direction opposite
to the one of the $z$-axis in the absence of the magnetic field.

The same dependence of the spin current on the magnetic field with small values of its magnitude (as well as for a small
value of the driving amplitude $FL=1.0\,\hbar\omega_0$) is found in Fig.~\ref{figure_2} in view of its dependence on the
spin-orbit interaction strength $k_\mathrm{so}$. Again for $H_0>0$ the magnitude of the spin-current gets smaller and for
$H_0<0$ it gets larger. Additionally, one can see that the presence of the magnetic field does not change the locations of
minima and maxima of the spin current as a function of $k_\mathrm{so}$. This has the following physical explanation. The
minima and maxima in Fig.~\ref{figure_2} are related to the periodicity of the energy spectrum in the $\vec{k}$-space. In
terms of the band energy versus the quasi-momentum $\vec{k}$ dependence RSOI produces a horizontal (that is the energy of
the bands does not change) split of the energy bands as well as their hybridization. Due to the periodicity this split can
be minimal or maximal for some values of $k_\mathrm{so}$ which leads to the corresponding minima and maxima in
Fig.~\ref{figure_2}. The role of the hybridization is that the split is never zero and thus the minima of the spin current
are not exactly equal to zero. In contrast to this horizontal split the magnetic field produces a vertical (that is along
the energy axis) split and it also produces hybridization. This vertical split is not correlated with the periodicity of
the energy bands in the $\vec{k}$-space and thus the locations of minima and maxima remain untouched by the magnetic
field.

\begin{figure}
\includegraphics[width=8.5 cm]{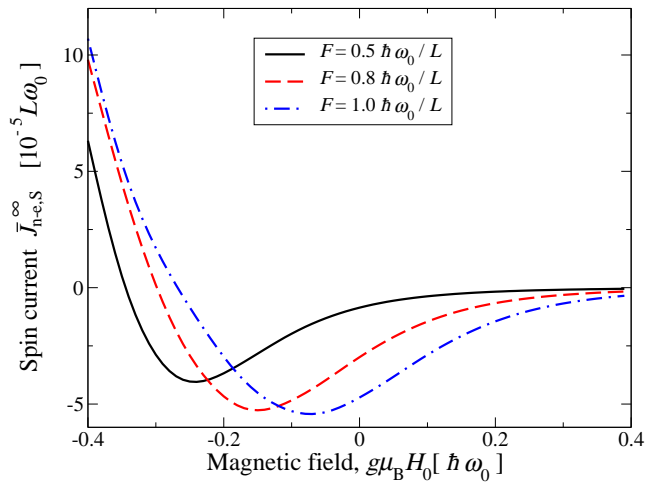}
\caption{\label{figure_3} (Color online) Non-equilibrium spin current, $\bar{J}^\infty_\mathrm{n-e,S}$, as a function of the
magnetic field, $g\mu_\mathrm{B}H_0$, for different values of the amplitude of the driving force, $F$. The other parameters
are as in Fig.~\ref{figure_1}.}
\end{figure}

However, the picture explained above is only valid for small values of the driving amplitude $F$ and magnitude of the
magnetic field $|H_0|$ where the spin current has a linear response to the magnetic field. When $|H_0|$ increases further,
the spin current depends non-linearly on $H_0$ and a complicated interplay between the magnetic field, driving and
dissipative processes develops. This dependence of the spin current on the magnetic field is depicted in
Fig.~\ref{figure_3} for different values of the amplitude of the driving force. In order to stay within the validity of
our model, where only the first two transverse modes are opened, the magnitude of the magnetic field must satisfy the
condition:
\begin{equation}
g\mu_\mathrm{B}|H_0|\leqslant 0.5(\hbar\omega_0+\Delta\varepsilon_{\gamma;4,1}),
\label{mag_fld_lim}
\end{equation}
where $\Delta\varepsilon_{\gamma;4,1}\equiv\sum_{k_\mathrm{B}}\bigl[\varepsilon_{\gamma,1;1}^{(0)}(k_\mathrm{B})-
\varepsilon_{\gamma,0;1}^{(0)}(k_\mathrm{B})\bigl]/N$. For the values of the parameters used to obtain the numerical results
we have $\Delta\varepsilon_{\gamma;4,1}=-0.07\hbar\omega_0$. Thus $g\mu_\mathrm{B}|H_0|\leqslant 0.465\hbar\omega_0$. As it
can be seen from Fig.~\ref{figure_3} the magnitude of the spin current decays for large positive values of $H_0$. This
happens because the distance between the Bloch sub-bands becomes large and thus the transition processes are less
probable. For a certain negative value of $H_0$ the magnitude of the spin current has a maximum after which it starts to
decrease and vanishes at some point $H_0^{(0)}<0$. After this point and for $H_0<H_0^{(0)}$ the spin current reverses its
sign and its magnitude increases again. This behavior clearly demonstrates that the magnetic field can, without changing
its direction, act in phase ({\it i.e.} destroy the spin transport) with the dissipative processes as well as out-of-phase
({\it i.e.} intensify the spin kinetics) with them. Mathematically it comes from the fact that in Eq.
(\ref{transition_rate}) for the transition rates the magnetic field $H_0$ and the imaginary part of the twice integrated
\begin{figure}
\includegraphics[width=8.5 cm]{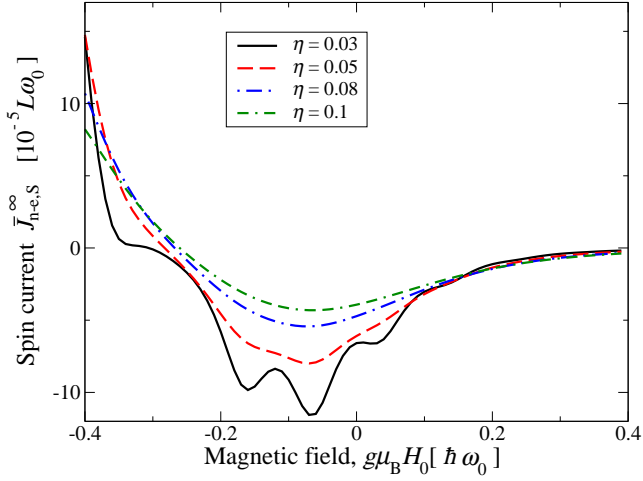}
\caption{\label{figure_4} (Color online) Non-equilibrium spin current, $\bar{J}^\infty_\mathrm{n-e,S}$, as a function of the
magnetic field, $g\mu_\mathrm{B}H_0$, for different values of the viscosity coefficient, $\eta$. The driving amplitude is
$F=1.0\hbar\omega_0/L$. The other parameters are as in Fig.~\ref{figure_1}.}
\end{figure}
bath correlation function $\mathrm{Im}[Q(\tau)]$ enter the arguments of the same trigonometric functions. This is
clarified by Eq. (\ref{integral_J}) appropriately rewritten below for the case $\xi'=1$, $\xi=4$:
\begin{equation}
\begin{split}
&J_{\gamma;1,4}=\frac{2}{\hbar^2}\int_0^\infty d\tau\mathrm{e}^{-\frac{L^2}{\hbar}Q_\mathrm{R}(\tau)}\times\\
&\times
\cos\biggl[\!\biggl(\frac{\Delta\varepsilon_{\gamma;4,1}}{\hbar}+\omega_0+\frac{2g\mu_\mathrm{B}H_0}{\hbar}\biggl)\tau-
\frac{L^2}{\hbar}Q_\mathrm{I}(\tau)\biggl]\times\\
&\times J_0\biggl[\frac{2FL}{\hbar\Omega}\sin\biggl(\frac{\Omega\tau}{2}\biggl)\biggl],
\end{split}
\label{integral_J_14}
\end{equation}
where $Q_\mathrm{R}(\tau)\equiv\mathrm{Re}[Q(\tau)]$, $Q_\mathrm{I}(\tau)\equiv\mathrm{Im}[Q(\tau)]$. The physical
explanation of why in our system the magnetic field interacts only with the friction part of the dissipation and not with
its noise part is rooted in the roles which the magnetic field and dissipation play for quantum coherence. On the one side
quantum coherence in a dissipative system dies out due to the noise part of the Feynman-Vernon influence weight
functional. On the other side, within the Feynman path integral formalism, we see that in our system a transverse in-plane
uniform stationary magnetic field cannot produce the additional phase due to the integral of the vector potential along
the Feynman paths (see Appendix \ref{energy_spinors_magnetic_field}). Thus in our system quantum coherence is totally
insensitive to the magnetic field and as a result cannot interact with the noise part of the Feynman-Vernon influence
weight functional.

The mutual impact of the magnetic field and quantum dissipative processes on the spin current in the system is shown in
Fig.~\ref{figure_4} where the spin current is plotted versus the magnetic field, $H_0$, and the viscosity coefficient,
$\eta$, plays a role of a parameter. Again for large positive values of $H_0$ the spin current vanishes. As expected, the
spin current gets smaller if the dissipation in the system gets stronger. When the dissipation gets weaker ($\eta=0.05$
and $\eta=0.03$ curves) the oscillations of the spin current become observable. These oscillations are related to the
interaction between the magnetic field and driving and can be described in terms of the photon emission/absorption
processes \cite{Grifoni_1} since changing $H_0$ is equivalent to changing the distance between the corresponding Bloch
sub-bands.

The minima in Figs.~\ref{figure_3} and \ref{figure_4} at negative values of $H_0$ appear as a result of a cooperative
action of the orbit-orbit coupling, confinement, magnetic field, driving and dissipation. Its location changes when the
strength of the driving and dissipation varies.

\begin{figure}
\includegraphics[width=8.0 cm]{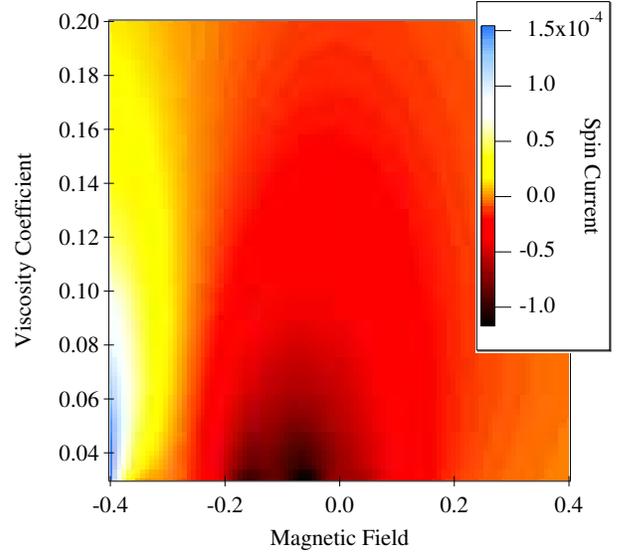}
\caption{\label{figure_5} (Color online) Contour plot of the non-equilibrium spin current, $\bar{J}^\infty_\mathrm{n-e,S}$
  [$L\omega_0$], as a function of the magnetic field, $g\mu_\mathrm{B}H_0$ [$\hbar\omega_0$], and viscosity coefficient,
  $\eta$. The other parameters are as in Fig.~\ref{figure_1}.}
\end{figure}
For completeness in Fig.~\ref{figure_5} we also show the spin current as a contour plot using the variables $H_0$ and
$\eta$. The main effect of the interaction between the electrons and external environment is the electron dressing. The
dressed electrons are heavier and as a result less mobile. Since the spin degree of freedom is carried by these dressed
electrons, the spin current decreases when the viscosity coefficient grows.

\section{Conclusion}\label{conclusion}
In conclusion, we have studied averaged stationary quantum transport in a driven dissipative periodic
quasi-one-dimensional (1D) system with Rashba spin-orbit interaction (RSOI) and placed in a transverse in-plane uniform
stationary magnetic field. For the case of moderate-to-strong dissipation it has been shown that the averaged stationary
charge transport is well suppressed as soon as it is restricted within the Bloch sub-bands grown out of the same Bloch
band of the corresponding truly 1D problem without RSOI. However in the same situation the averaged stationary spin
transport is activated. The analytical expression for the spin current has been derived and its behavior as a function of
the driving parameters, dissipation, spin-orbit interaction strength, orbit-orbit coupling strength and a transverse
in-plane uniform stationary magnetic field has been analyzed. Our results on the spin transport in the system have been
presented and thoroughly discussed. It has been found that the spin current as a function of the magnetic field shows a
highly non-trivial dependence for different values of the dissipation and driving parameters. In particular, increasing
the magnitude of the magnetic field does not always lead to a monotonous response in the magnitude of the spin current.
The magnitude of the spin current can have maxima after which its dependence on the magnitude of the magnetic field
changes to the opposite one. Moreover, the spin current as a function of the amplitude of an external longitudinal ac
electric field has reversals of its direction when the system is placed in a finite transverse in-plane uniform stationary
magnetic field. Also as a function of this magnetic field the spin current changes its direction at finite values of the
amplitude of the ac electric field. Such behavior is undoubtedly related to a deep correlation between the dissipative
processes and magnetic field effects in the system.

\begin{acknowledgments}
Support from the DFG under the program SFB 689 is acknowledged.
\end{acknowledgments}

\appendix
\section{Eigen-value structure of the coordinate operator in a subspace generated by Bloch states of a
finite number of bands}\label{eigen_value_x}
In this appendix we consider a physical property with the corresponding quantum mechanical operator which when operating
in the Hilbert space has a continuum spectrum and show how this continuum spectrum can turn into a discrete one under a
certain restriction of the Hilbert space. To be specific we constrict the Hilbert space to a subspace using some of the
Bloch states and consider how the coordinate operator changes its spectrum.

\subsection{Introduction}
In many problems of condensed matter theory one is not usually interested in the full band structure of a solid but rather
in a few bands most important for the relevant physics of a system. For example in metals one or a few bands with an
energy range containing the Fermi energy are most important since the main contributions to transport properties come
almost only from those bands. Taking into account only a few Bloch bands leads to a restriction of the Hilbert space to a
subspace which is then used to describe physical properties.

\subsection{Truncation of the Hilbert space using Bloch states}
Let $\mathcal{H}$ be the Hilbert space of all possible states and let us choose in this space the basis of Bloch's states
$\{|l,k_\mathrm{B}\rangle\}$:
\begin{equation}
\begin{split}
&\langle x|l,k_\mathrm{B}\rangle=e^{\mathrm{i}k_\mathrm{B}x}u_{l,k_\mathrm{B}}(x),\\
&u_{l,k_\mathrm{B}}(x+L)=u_{l,k_\mathrm{B}}(x),\\
&\forall\,\, k_\mathrm{B}\in\mathrm{B.Z.},\, l=1,2,\ldots\,,
\end{split}
\label{Bloch_basis}
\end{equation}
where $L$ is the period of the Bloch amplitude $u_{l,k_\mathrm{B}}(x)$ and $\mathrm{B.Z.}$ stands for the first Brillouin
zone.

Any vector $|\psi\rangle\in\mathcal{H}$ represents a linear combination
\begin{equation}
|\psi\rangle=\sum_{l=1}^{\infty}\sum_{k_\mathrm{B}\in\mathrm{B.Z.}}c_{l,k_\mathrm{B}}|l,k_\mathrm{B}\rangle.
\label{linear_combination_full}
\end{equation}

Another basis $|\alpha\rangle$ is obtained using a transformation
\begin{equation}
|\alpha\rangle=\hat{U}^{-1}|l,k_\mathrm{B}\rangle,\quad \forall\,\, k_\mathrm{B}\in\mathrm{B.Z.},\, l=1,2,\ldots\,,
\label{another_basis}
\end{equation}
where $\hat{U}$ is an arbitrary unitary operator.

Let us consider an operator $\hat{\mathcal{O}}$ corresponding to an observable $\mathcal{O}$. Its matrix representations
in the two bases (\ref{Bloch_basis}) and (\ref{another_basis}) are
\begin{equation}
\begin{split}
&\mathcal{O}_\mathrm{B}=\langle l',k_\mathrm{B}'|\hat{\mathcal{O}}|l,k_\mathrm{B}\rangle,\\
&\forall\,\, k_\mathrm{B},k_\mathrm{B}'\in\mathrm{B.Z.},\, l,l'=1,2,\ldots\,,\\
&\mathcal{O}_\alpha=\langle\alpha'|\hat{\mathcal{O}}|\alpha\rangle,\quad \forall\,\, \alpha,\alpha'.
\end{split}
\label{full_matrices}
\end{equation}
The eigen-values $\{\lambda_i\}$ of the two matrices (\ref{full_matrices}) are the same and represent all possible values
of the observable $\mathcal{O}$.

Now let us consider a subspace $\mathcal{S}\subset\mathcal{H}$ generated by Bloch's states corresponding to a finite
number, $N_\mathrm{B}$, of bands. A vector $|\psi^\mathcal{S}\rangle\in\mathcal{S}$ has the form:
\begin{equation}
|\psi\rangle=\sum_{i=1}^{N_\mathrm{B}}\sum_{k_\mathrm{B}\in\mathrm{B.Z.}}c_{l_i,k_\mathrm{B}}|l_i,k_\mathrm{B}\rangle.
\label{linear_combination}
\end{equation}
In this subspace the operator $\hat{\mathcal{O}}$ has the matrix representation:
\begin{equation}
\begin{split}
&\mathcal{O}_\mathrm{B}^{\mathcal{S}}=\langle l_{i'},k_\mathrm{B}'|\hat{\mathcal{O}}|l_i,k_\mathrm{B}\rangle,\\
&\forall\,\, k_\mathrm{B},k_\mathrm{B}'\in\mathrm{B.Z.},
\, i,i'=1,2,\ldots N_\mathrm{B}.
\end{split}
\label{matrix_Bloch_cut}
\end{equation}
Now the eigen-values $\{\lambda_n^{\mathcal{S}}\}$ of (\ref{matrix_Bloch_cut}) do not represent all possible values of the
observable $\mathcal{O}$ but they only give approximate values of some of them. If the operator $\hat{\mathcal{O}}$
corresponds to a continuous observable with the spectrum from $-\infty$ to $\infty$, the eigen-values
$\{\lambda_n^{\mathcal{S}}\}$ are some of the eigen-values $\{\lambda_i\}$, that is in this case
$\{\lambda_n^{\mathcal{S}}\}\subset\{\lambda_i\}$.

A new basis $\{\alpha^\mathcal{S}\}$ of the subspace $\mathcal{S}$ is related to the Bloch one as:
\begin{equation}
\begin{split}
&|\alpha^\mathcal{S}\rangle=\hat{U}_\mathcal{S}^{-1}|l_i,k_\mathrm{B}\rangle,\\
&\forall\,\, k_\mathrm{B}\in\mathrm{B.Z.},\, i=1,2,\ldots,N_\mathrm{B},
\end{split}
\label{another_basis_cut}
\end{equation}
where now $\hat{U}_\mathcal{S}^{-1}$ is not an arbitrary unitary operator, but a unitary operator with the following
property:
\begin{equation}
\hat{U}_{\mathcal{S}}:\quad |v\rangle\in\mathcal{S}\,\,\Rightarrow\,\,\hat{U}_{\mathcal{S}}|v\rangle\in\mathcal{S},
\quad \forall\,\, |v\rangle\in\mathcal{S}.
\label{unitary_cut}
\end{equation}
In this case the matrix
\begin{equation}
\mathcal{O}_{\alpha}^{\mathcal{S}}=\langle\alpha'^{\mathcal{S}}|\hat{\mathcal{O}}|\alpha^{\mathcal{S}}\rangle,
\quad \forall\,\,\alpha'^{\mathcal{S}},\alpha^{\mathcal{S}}
\label{matrix_another_cut}
\end{equation}
has the same set of eigen-values $\{\lambda_n^{\mathcal{S}}\}$ as the matrix $\mathcal{O}_\mathrm{B}^{\mathcal{S}}$ in
(\ref{matrix_Bloch_cut}).

\subsection{Example: coordinate}
Let us specify the observable $\mathcal{O}$ from the preceding section to be particle's coordinate $q$ with the
corresponding operator denoted as $\hat{q}$. We consider the operator $\hat{q}$ in the subspace $\mathcal{S}$. Its matrix
with respect to the Bloch basis is
\begin{equation}
\begin{split}
&q_\mathrm{B}^{\mathcal{S}}=\langle l_{i'},k_\mathrm{B}'|\hat{q}|l_i,k_\mathrm{B}\rangle,\\
&\forall\,\, k_\mathrm{B},k_\mathrm{B}'\in\mathrm{B.Z.},\, i,i'=1,2,\ldots,N_\mathrm{B}.
\end{split}
\label{q_matrix_cut}
\end{equation}

Let us choose the translational operator as the unitary operator $\hat{U}_{\mathcal{S}}$ from the preceding section, that is
\begin{equation}
\hat{U}_{\mathcal{S}}(a)=e^{\frac{i}{\hbar}a\hat{p}}.
\label{translational_op}
\end{equation}
It is obvious that for an arbitrary value of $a$ the operator $\hat{U}_{\mathcal{S}}(a)$ does not satisfy the property
(\ref{unitary_cut}). However, in the case $a=L$ a Bloch state $|l,k_\mathrm{B}\rangle$ is translated into a Bloch state
with the same $l$, $k_\mathrm{B}$ and thus (\ref{unitary_cut}) is fulfilled. Hence, the matrix
\begin{equation}
\begin{split}
&\tilde{q}_\mathrm{B}^{\mathcal{S}}=
\langle l_{i'},k_\mathrm{B}'|\hat{U}_{\mathcal{S}}(L)\hat{q}\hat{U}_{\mathcal{S}}^{-1}(L)|l_i,k_\mathrm{B}\rangle,\\
&\forall\,\, k_\mathrm{B},k_\mathrm{B}'\in\mathrm{B.Z.},\, i,i'=1,2,\ldots,N_\mathrm{B}
\end{split}
\label{q_tld_matrix_cut}
\end{equation}
has the same eigen-values as the matrix $q_\mathrm{B}^{\mathcal{S}}$ in (\ref{q_matrix_cut}). But due to the equality
\begin{equation}
\hat{U}_{\mathcal{S}}(L)\hat{q}\hat{U}_{\mathcal{S}}^{-1}(L)=\hat{q}+L
\end{equation}
the two matrices $q_\mathrm{B}^{\mathcal{S}}$ and $\tilde{q}_\mathrm{B}^{\mathcal{S}}$ are related as follows:
\begin{equation}
\begin{split}
&\langle l_{i'},k_\mathrm{B}'|\hat{U}_{\mathcal{S}}(L)\hat{q}\hat{U}_{\mathcal{S}}^{-1}(L)|l_i,k_\mathrm{B}\rangle=\\
&=\langle l_{i'},k_\mathrm{B}'|\hat{q}|l_i,k_\mathrm{B}\rangle+L\delta_{i',i}\delta_{k_\mathrm{B}',k_\mathrm{B}},\\
&\forall\,\, k_\mathrm{B},k_\mathrm{B}'\in\mathrm{B.Z.},\, i,i'=1,2,\ldots,N_\mathrm{B}.
\end{split}
\label{relation_q_q_tld}
\end{equation}
From (\ref{relation_q_q_tld}) it follows that the eigen-values of the matrix $q_\mathrm{B}^{\mathcal{S}}$ are invariant
under a shift equal to $jL$ with $j$ being an integer. That is for any
$\lambda_{k}^{\mathcal{S}}\in\{\lambda_n^\mathcal{S}\}$ there exists $\lambda_{m}^{\mathcal{S}}\in\{\lambda_n^\mathcal{S}\}$
such that
\begin{equation}
\lambda_{k}^{\mathcal{S}}=jL+\lambda_{m}^{\mathcal{S}}.
\end{equation}
Let us denote through $\{d_k^\mathcal{S}\}$ those eigen-values of $q_\mathrm{B}^{\mathcal{S}}$ the distance between which is
less than $L$,
\begin{equation}
|d_r^\mathcal{S}-d_{r'}^\mathcal{S}|<L,\quad \forall\,\, d_r^\mathcal{S},d_{r'}^\mathcal{S}\in\{d_k^\mathcal{S}\},
\end{equation}
and which are in the zeroth elementary cell. Then each of the eigen-values $\{\lambda_n^\mathcal{S}\}$ of the matrix
$q_\mathrm{B}^{\mathcal{S}}$ is obtained from its corresponding eigen-value $d_m\in\{d_k^\mathcal{S}\}$ by a shift $jL$ with
a proper integer $j$. It means that each elementary cell contains the same number of eigen-values of the coordinate
operator. Since the total number of the eigen-values $\{\lambda_n^\mathcal{S}\}$ is equal to $N_\mathrm{B}N$ where $N$ is
the number of the elementary cells, we conclude that there are $N_\mathrm{B}$ eigen-values in each elementary cell. This
gives us the final expression for the eigen-values of the matrix $q_\mathrm{B}^{\mathcal{S}}$ ($N$ is even to be definite):
\begin{equation}
\begin{split}
&\lambda_{m,j}^\mathcal{S}=jL+d_m^\mathcal{S}, m=1,2,\ldots,N_\mathrm{B},\\
&j\!=\!-\frac{N}{2}\!+\!1,-\frac{N}{2}+2,\ldots,-1,0,1,\ldots,\frac{N}{2}-1,\frac{N}{2}
\end{split}
\end{equation}
and $N\to\infty$ afterwards.

\section{Polarization of the spin current along the $x$ and $y$ directions. Conventional spin current}
\label{spin_current_x_y_conv_SC}
In Ref.~\onlinecite{Smirnov_1} it was shown that in an isolated system (without any external magnetic field) the only
non-vanishing spin polarization is along the confinement direction, that is along the $z$-axis. A natural question is then
what is going on in an open driven system in a uniform stationary magnetic field applied along the $z$-axis. The external
force (\ref{driving_hamiltonian}) and bath Hamiltonian (\ref{bath_hamiltonian}) couple to the longitudinal orbital degree
of freedom, that is to the $x$-coordinate of our system. Because of the spin-orbit coupling the external force and bath
affect the spin dynamics of electrons in the quasi-1D system. The magnetic field also influences the spin transport. Can
it then happen that the longitudinal spin current acquires components polarized along the $x$ and $y$ axes? Below we show
that the components $J_\mathrm{S}^{x,y}(t)$ of the spin current (\ref{spin_current}) identically vanish.

\subsection{Longitudinal spin current components $J_\mathrm{S}^{x,y}(t)$ polarized along the $x$ and $y$ axes}
The expressions for the spin currents
\begin{equation}
J_\mathrm{S}^{x,y}(t)=\frac{d}{dt}\mathrm{Tr}\bigl(\hat{\sigma}_{x,y}\hat{x}\hat{\rho}(t)\bigl)
\label{spin_currents_x_y}
\end{equation}
can easily be found using the $\sigma$-DVR basis $\{|\zeta,m,j,\sigma\rangle_{\gamma,j}\}$ introduced in Section
\ref{Diag_sigma_x}:
\begin{equation}
\begin{split}
&J_\mathrm{S}^{x}(t)=2\frac{d}{dt}\mathrm{Tr}_\mathrm{B}\sum_{\zeta,m,j}(mL+d_{\gamma;\zeta,j})\times\\
&\times\mathrm{Re}\bigl(\, _{\gamma,j}\langle\zeta,m,j,\sigma'=+1|\hat{W}(t)|\zeta,m,j,\sigma=-1\rangle_{\gamma,j}\bigl),
\end{split}
\label{spin_current_sigma_DVR_x}
\end{equation}
\begin{equation}
\begin{split}
&J_\mathrm{S}^{y}(t)=-2\frac{d}{dt}\mathrm{Tr}_\mathrm{B}\sum_{\zeta,m,j}(mL+d_{\gamma;\zeta,j})\times\\
&\times\mathrm{Im}\bigl(\, _{\gamma,j}\langle\zeta,m,j,\sigma'=+1|\hat{W}(t)|\zeta,m,j,\sigma=-1\rangle_{\gamma,j}\bigl),
\end{split}
\label{spin_current_sigma_DVR_y}
\end{equation}
where we have explicitly written the trace over the bath degrees of freedom in order to work further with the $\sigma$-DVR
matrix elements of the full statistical operator $\hat{W}(t)$.

\subsection{Selection rules for the $\sigma$-DVR matrix elements of the full statistical operator}
It turns out that the case of a harmonic confinement allows one to formulate selection rules for the $\sigma$-DVR matrix
elements of the full statistical operator. These selection rules are very useful for understanding some of the properties
of the spin transport.

To find the selection rules mentioned above let us decompose the Hamiltonian $\hat{H}$ in (\ref{isolated_hamiltonian})
into
\begin{equation}
\hat{H}=\hat{H}_0+\hat{H}_\mathrm{R-Z},
\label{isolated_hamiltonian_H0_HR}
\end{equation}
where
\begin{equation}
\hat{H_0}=\frac{\hbar^2\hat{\vec{k}}^2}{2m}+\frac{m\omega_0^2\hat{z}^2}{2}+U(\hat{x})
\biggl(1+\gamma\frac{\hat{z}^2}{L^2}\biggl),
\label{hamiltonian_H0}
\end{equation}
\begin{equation}
\begin{split}
\hat{H}_\mathrm{R-Z}&=-\frac{\hbar^2k_{\mathrm{so}}}{m}\bigl(\hat{\sigma}_{x}\hat{k}_z-\hat{\sigma}_z\hat{k}_x\bigl)-
g\mu_\mathrm{B}\hat{\sigma}_zH_0=\\
&=-\frac{\hbar^2k_{\mathrm{so}}}{m}\bigl(\hat{\sigma}_{x}\hat{k}_z-\hat{\sigma}_z\hat{k}_x'\bigl),
\end{split}
\label{hamiltonian_HR}
\end{equation}
and $\hat{k}_x'=\hat{k}_x-g\mu_\mathrm{B}H_0m/\hbar^2 k_\mathrm{so}$. The full statistical operator has the form
$\hat{W}(t)=\hat{U}(t,t_0)\hat{W}(t_0)\hat{U}^\dagger(t,t_0)$, where the evolution operator $\hat{U}(t,t_0)$ is given as
the time-ordered exponent
\begin{equation}
\begin{split}
&\hat{U}(t,t_0)=\mathrm{T}\exp\biggl[-\frac{\mathrm{i}}{\hbar}\int_{t_0}^tdt'\hat{H}_\mathrm{full}(t')\biggl]=\\
&=\sum_{n=0}^\infty\biggl(-\frac{\mathrm{i}}{\hbar}\biggl)^n\!\!\!\int_{t_0}^t\!\!\!\!dt_n\cdots\int_{t_0}^{t_2}\!\!\!\!dt_1
\hat{H}_\mathrm{full}(t_n)\cdots\hat{H}_\mathrm{full}(t_1).
\end{split}
\label{evol_op}
\end{equation}
Only the terms of $\hat{H}_\mathrm{R-Z}$ with odd powers contain the spin operators. These terms are linear in
$\hat{\sigma}_x$ and $\hat{\sigma}_z$ or bilinear in these spin operators which is equivalent to being linear in
$\hat{\sigma}_y$. Contributions to the matrix elements
$_{\gamma,j}\langle\zeta,m,j,\sigma'=+1|\hat{W}(t)|\zeta,m,j,\sigma=-1\rangle_{\gamma,j}$ come from the first order terms in
$\hat{\sigma}_x$. It is easy to see that these terms represent products of the factors
$(\hat{H}_0+\hat{H}_\mathrm{ext}(t_k)+\hat{H}_\mathrm{bath})$ ordered chronologically (we mean the chronological ordering on
the Keldysh contour \cite{Keldysh} and thus operators from $\hat{U}^\dagger(t,t_0)$ are also included under this
terminology), an odd number of factors $\hat{k}_z$ distributed in between
$(\hat{H}_0+\hat{H}_\mathrm{ext}(t_k)+\hat{H}_\mathrm{bath})$ in all possible ways and a number (even or odd) of factors
$\hat{k}_x'$ also distributed in between $(\hat{H}_0+\hat{H}_\mathrm{ext}(t_k)+\hat{H}_\mathrm{bath})$ in all possible ways.
Such a structure is related to the fact that the Rashba-Zeeman Hamiltonian, $\hat{H}_\mathrm{R-Z}$, is bilinear in the
operators $\hat{\sigma}_x$ and $\hat{k}_z$. To clarify our above statement we write down the third order term coming for
example from $\hat{U}(t,t_0)$ (a similar result is obtained for products which are composed from different,
$\hat{U}(t,t_0)$, $\hat{U}^\dagger(t,t_0)$ or $\hat{W}(t_0)$, parts of the full statistical operator):
\begin{equation}
\begin{split}
&\hat{H}_\mathrm{full}(t_3)\hat{H}_\mathrm{full}(t_2)\hat{H}_\mathrm{full}(t_1)=\hat{H}_\mathrm{R-Z}^3+
\hat{H}_\mathrm{R-Z}^2\bigl(\hat{H}_0+\\
&+\hat{H}_\mathrm{ext}(t_1)+\hat{H}_\mathrm{bath}\bigl)+\hat{H}_\mathrm{R-Z}\bigl(\hat{H}_0+\hat{H}_\mathrm{ext}(t_2)+\\
&+\hat{H}_\mathrm{bath}\bigl)\hat{H}_\mathrm{R-Z}\!+\!\hat{H}_\mathrm{R-Z}(\hat{H}_0\!+\!\hat{H}_\mathrm{ext}(t_2)+
\hat{H}_\mathrm{bath}\bigl)\times\\
&\times(\hat{H}_0+\hat{H}_\mathrm{ext}(t_1)+\hat{H}_\mathrm{bath}\bigl)+
(\hat{H}_0+\hat{H}_\mathrm{ext}(t_3)+\\
&+\hat{H}_\mathrm{bath}\bigl)\hat{H}_\mathrm{R-Z}^2\!+\!(\hat{H}_0+\hat{H}_\mathrm{ext}(t_3)\!+\!\hat{H}_\mathrm{bath}\bigl)
\hat{H}_\mathrm{R-Z}\times\\
&\times(\hat{H}_0+\hat{H}_\mathrm{ext}(t_1)+\hat{H}_\mathrm{bath}\bigl)+(\hat{H}_0+\hat{H}_\mathrm{ext}(t_3)+\\
&+\hat{H}_\mathrm{bath}\bigl)(\hat{H}_0+\hat{H}_\mathrm{ext}(t_2)+\hat{H}_\mathrm{bath}\bigl)\hat{H}_\mathrm{R-Z}+
(\hat{H}_0+\\
&+\hat{H}_\mathrm{ext}(t_3)\!+\!\hat{H}_\mathrm{bath}\bigl)(\hat{H}_0+\hat{H}_\mathrm{ext}(t_2)+\hat{H}_\mathrm{bath}\bigl)
(\hat{H}_0+\\
&+\hat{H}_\mathrm{ext}(t_1)+\hat{H}_\mathrm{bath}\bigl).
\end{split}
\label{third_order_evol_op}
\end{equation}

Since for a harmonic confinement all the factors $(\hat{H}_0+\hat{H}_\mathrm{ext}(t_k)+\hat{H}_\mathrm{bath})$ and
$\hat{k}_x'$ couple states with indices $j$ and $j'$ only of identical parity and the factors $\hat{k}_z^{2m+1}$ couple
states with indices $j$ and $j'$ only of opposite parity, we conclude that the matrix elements
$_{\gamma,j}\langle\zeta,m,j,\sigma'=+1|\hat{W}(t)|\zeta,m,j,\sigma=-1\rangle_{\gamma,j}$, being diagonal in $j$, are equal
to zero:
\begin{equation}
_{\gamma,j}\langle\zeta,m,j,\sigma'=+1|\hat{W}(t)|\zeta,m,j,\sigma=-1\rangle_{\gamma,j}=0.
\label{selection_rules}
\end{equation}
The selection rules (\ref{selection_rules}) represent a specific property of systems with a harmonic confinement. From
(\ref{selection_rules}) one gets
\begin{equation}
J_\mathrm{S}^{x,y}(t)=0.
\end{equation}
In spite of the fact that this result is only valid for the case of a harmonic confinement it is still general in two
respects: 1) it is valid not only for the stationary state but for all times $t\geqslant t_0$; 2) the external force $F(t)$
is arbitrary.

\subsection{Role of the spin current definition}
In light of the mathematical formalism of the this appendix it is now convenient to discuss the difference between the
conventional spin current definition and the definition of the spin current used in our work, that is the definition
introduced by Shi {\it et al.}\cite{Shi}. We will consider the $z$-polarized components of the spin currents obtained from
the two definitions. The conventional spin current operator and the conventional spin current will be denoted as
$\hat{J}_\mathrm{S}^\mathrm{conv}(t)$ and $J_\mathrm{S}^\mathrm{conv}(t)$, respectively. The spin current operator and the
spin current which are used in our work will be denoted as $\hat{J}_\mathrm{S}(t)$ and $J_\mathrm{S}(t)$, respectively.

The two definitions and the difference between them are
\begin{equation}
\begin{split}
&J_\mathrm{S}(t)=\frac{d}{dt}(\hat{\sigma}_z\hat{x}),\quad
J_\mathrm{S}^\mathrm{conv}(t)=\hat{\sigma}_z\frac{d\hat{x}}{dt},\\
&J_\mathrm{S}(t)-J_\mathrm{S}^\mathrm{conv}(t)=\frac{d\hat{\sigma}_z}{dt}\hat{x}.
\end{split}
\label{SC_shi_conv_diff}
\end{equation}
One easily finds that
\begin{equation}
\frac{d\hat{\sigma}_z}{dt}=-\frac{2\hbar k_\mathrm{so}}{m}\hat{\sigma}_y\hat{k}_z.
\label{d_sigma_z_d_t}
\end{equation}
Thus the relation between the spin currents is
\begin{equation}
\begin{split}
&J_\mathrm{S}^\mathrm{conv}(t)=J_\mathrm{S}(t)+\\
&+\mathrm{i}\frac{2\hbar k_\mathrm{so}}{m}\mathrm{Tr}_\mathrm{B}\sum_{\zeta,m,j}(mL+d_{\gamma;\zeta,j})\times\\
&\times\bigl(\, _{\gamma,j}\langle\zeta,m,j,\sigma'=+1|\hat{k}_z\hat{W}(t)|\zeta,m,j,\sigma=-1\rangle_{\gamma,j}-\\
&-\, _{\gamma,j}\langle\zeta,m,j,\sigma'=-1|\hat{k}_z\hat{W}(t)|\zeta,m,j,\sigma=+1\rangle_{\gamma,j}\bigl).
\end{split}
\label{SC_relation}
\end{equation}
The second term in Eq. (\ref{SC_relation}) can be finite for our system. To show this we consider the product
$\hat{H}_\mathrm{full}(t_3)\hat{H}_\mathrm{full}(t_2)\hat{H}_\mathrm{full}(t_1)$ in Eq. (\ref{third_order_evol_op}). This
product contains for example the term $\hat{H}^3$ where $\hat{H}$ is given by Eq. (\ref{isolated_hamiltonian_H0_HR}). We
can write this term as
\begin{equation}
\begin{split}
&\hat{H}^3=\hat{H}_0^3+\biggl(\frac{\hbar^2 k_\mathrm{so}}{m}\biggl)^2\hat{H}_0\hat{\vec{k}}^2-
\frac{\hbar^2 k_\mathrm{so}}{m}\bigl[\hat{H}_0^2(\hat{\sigma}_x \hat{k}_z-\\
&-\hat{\sigma}_z \hat{k}_x')+\hat{H}_0(\hat{\sigma}_x \hat{k}_z-\hat{\sigma}_z \hat{k}_x')\hat{H}_0\bigl]-
\frac{\hbar^2 k_\mathrm{so}}{m}(\hat{\sigma}_x \hat{k}_z-\\
&-\hat{\sigma}_z \hat{k}_x')\hat{H}_0^2-\biggl(\frac{\hbar^2 k_\mathrm{so}}{m}\biggl)^3(\hat{\sigma}_x \hat{k}_z-
\hat{\sigma}_z \hat{k}_x')\hat{\vec{k}}^2+\\
&+\biggl(\frac{\hbar^2 k_\mathrm{so}}{m}\biggl)^2\bigl[\hat{k}_z\hat{H}_0\hat{k}_z+\hat{k}_x'\hat{H}_0\hat{k}_x'+
\mathrm{i}\hat{\sigma}_y(\hat{k}_z\hat{H}_0\hat{k}_x'-\\
&-\hat{k}_x'\hat{H}_0\hat{k}_z)+\hat{\vec{k}}^2\hat{H}_0^2\bigl].
\end{split}
\label{H_cube}
\end{equation}
From Eq. (\ref{H_cube}) we see that the operator $\hat{k}_z\hat{W}(t)$ has terms like
\begin{equation}
\mathrm{i}\biggl(\frac{\hbar^2 k_\mathrm{so}}{m}\biggl)^2\hat{k}_z\hat{\sigma}_y(\hat{k}_z\hat{H}_0\hat{k}_x'-
\hat{k}_x'\hat{H}_0\hat{k}_z),
\label{sigma_y_even_k_z}
\end{equation}
which are even with respect to $\hat{k}_z$ and odd with respect to $\hat{\sigma}_y$. Therefore, in general we have
\begin{equation}
\begin{split}
&\, _{\gamma,j}\langle\zeta,m,j,\sigma'=+1|\hat{k}_z\hat{W}(t)|\zeta,m,j,\sigma=-1\rangle_{\gamma,j}-\\
&-\, _{\gamma,j}\langle\zeta,m,j,\sigma'=-1|\hat{k}_z\hat{W}(t)|\zeta,m,j,\sigma=+1\rangle_{\gamma,j}\neq 0,
\end{split}
\label{non_symmetry}
\end{equation}
which means that the two spin current definitions are different in our problem. The physical reason for this can be
understood from Eq. (\ref{sigma_y_even_k_z}). The term given by Eq. (\ref{sigma_y_even_k_z}) is finite since $\hat{k}_x'$
and $\hat{k}_z$ do not commute with $\hat{H}_0$. It happens because of the presence of both the periodic potential and the
confinement as it is obvious from Eq. (\ref{hamiltonian_H0}). Thus we conclude that unlike free Rashba electrons the two
definitions of the spin current are not equivalent for Rashba-Bloch electrons with a transverse confinement.

As one can see from Eqs. (\ref{averaged_currents}), (\ref{transition_rate}) and
(\ref{averaged_master_equation_full_sigma_DVR_tb}) in the insulating limit $\bar{J}_\mathrm{S}(t)\rightarrow 0$. This is
just a consequence of the fact that the spin current definition which we use represents a full derivative. It is quite
reasonable from the physical point of view that the spin ratchet effect being a transport phenomenon is absent in
insulators. However, the conventional definition of the spin current is not a full derivative. The spin current
$J_\mathrm{S}^\mathrm{conv}(t)$ differs from the spin current $J_\mathrm{S}(t)$ by the second term in Eq.
(\ref{SC_relation}). There is not any general physical reason for this term, averaged over one driving period, to vanish
in the insulating limit at long times because it is not proportional to the time derivative of the averaged populations of
the states but it is proportional to the averaged non-diagonal (in the spin and transverse mode subspaces) elements of the
reduced statistical operator. These averaged non-diagonal elements can in general be finite in insulators. The spin
ratchet effect obtained from the conventional definition of the spin current could then take place in insulators which to
our opinion would be unphysical.

\section{Eigen-energies and eigen-spinors in the presence of orbit-orbit coupling and a uniform stationary
magnetic field along the $z$-axis}
\label{energy_spinors_magnetic_field}
In Ref.~\onlinecite{Smirnov_1} periodic structures formed in a 2DEG with RSOI have been considered. However the influence
of an external homogeneous stationary magnetic field on the energy spectrum has not been studied. Here we generalize the
results of Ref.~\onlinecite{Smirnov_1} to the case of a uniform stationary magnetic field applied along the $z$-axis.
Afterwards we discuss the orbit-orbit coupling introduced in the main text in Eq. (\ref{isolated_hamiltonian}).

\subsection{System's Hamiltonian and the general eigen-value equation}
For an arbitrary potential $V(z)$ (not necessarily confinement) along the $z$-axis and a uniform stationary magnetic field
applied along the $z$-axis (2DEG is in the $x-z$ plane) the Hamiltonian reads
\begin{equation}
\begin{split}
\hat{H}=\frac{\hbar^2\hat{\vec{k}}^2}{2m}+V(\hat{z})&-
\frac{\hbar^2k_{\mathrm{so}}}{m}\bigl(\hat{\sigma}_x\hat{k}_z-\hat{\sigma}_z\hat{k}_x\bigl)+\\
&+U(\hat{x})-g\mu_\mathrm{B}\hat{\sigma}_zH_0.
\end{split}
\label{hamiltonian_general}
\end{equation}
In Eq. (\ref{hamiltonian_general}) $H_0$ is the $z$-component of the magnetic field $\vec{H}=(0,0,H_0)$ and the Landau
gauge, $\vec{A}=(-H_0y,0,0)$, has been chosen. Additionally we have used the fact that in a 2DEG $y=0$. This choice
effectively gives only the Zeeman term. The eigen-states of Hamiltonian (\ref{hamiltonian_general}) are Bloch spinors with
the spinorial amplitude given as (see Ref.~\onlinecite{Smirnov_1})
\begin{equation}
u_{l,k_\mathrm{B},\eta}(x;j,\sigma)=u_{l,k_\mathrm{B}+\sigma k_\mathrm{so}}(x)\theta_{l,k_\mathrm{B},\eta}(j,\sigma),
\label{spinorial_amplitude}
\end{equation}
where $u_{l,k_\mathrm{B}}(x)$ is the Bloch amplitude of the corresponding truly 1D problem without the magnetic field and
without RSOI, and $\theta_{l,k_\mathrm{B},\eta}(j,\sigma)$ is the eigen-spinor. This eigen-spinor is obtained from the
solution of the eigen-value equation for Hamiltonian (\ref{hamiltonian_general}):
\begin{equation}
\begin{split}
&\sum_{j',\sigma'}\biggl\{\delta_{j,j'}\delta_{\sigma,\sigma'}\biggl[\varepsilon_l^{(0)}(k_\mathrm{B}+\sigma k_\mathrm{so})-
g\mu_\mathrm{B}\sigma H_0+\varepsilon_j^z-\\
&-\!\frac{\hbar^2 k_\mathrm{so}^2}{2m}\biggl]\!-\frac{\hbar^2 k_\mathrm{so}}{m}(1\!-\!\delta_{\sigma,\sigma'})
\langle j|\hat{k}_z|j'\rangle\!\!\biggl\}\theta_{l,k_\mathrm{B},\eta}(j',\sigma')\!=\\
&=\varepsilon_{l,\eta}(k_\mathrm{B})\theta_{l,k_\mathrm{B},\eta}(j,\sigma).
\end{split}
\label{eigen_value_eq_general}
\end{equation}

\subsection{Harmonic confinement}
The case of a harmonic confinement is characterized by the following matrix elements of the operator $\hat{k}_z$:
\begin{equation}
\langle j|\hat{k}_z|j'\rangle=\pm\mathrm{i}\delta_{j,j'\pm 1}
\sqrt{\frac{\bigl(j+\frac{1}{2}\mp\frac{1}{2}\bigl)m\omega_0}{2\hbar}},
\label{matrix_elements_k_z}
\end{equation}
and eigen-energies $\varepsilon_j^z$:
\begin{equation}
\varepsilon_j^z=\hbar\omega_0\biggl(j+\frac{1}{2}\biggl).
\label{eigen_energies_epsilon_j_z}
\end{equation}
Therefore the only change in comparison with Ref.~\onlinecite{Smirnov_1} is in the diagonal matrix elements of the
Hamiltonian. Reproducing the same calculations as in Ref.~\onlinecite{Smirnov_1}, that is taking into account only the
first two transverse modes ($j=0,1$, $\sigma=\pm1$, $\eta=1,2,3,4$), one finds that the only change in the final results
for the eigen-energies and eigen-spinors consists in replacing the function $\varepsilon_l^-(k_\mathrm{B})$ with
\begin{equation}
\begin{split}
\varepsilon_l^-(k_\mathrm{B};H_0)=&
\frac{\varepsilon_l^{(0)}(k_\mathrm{B}+k_\mathrm{so})-\varepsilon_l^{(0)}(k_\mathrm{B}-k_\mathrm{so})}{2}-\\
&-g\mu_\mathrm{B}H_0,
\end{split}
\label{epsilon_minus}
\end{equation}
where we have explicitly shown the dependence on the $z$-component $H_0$ of the magnetic field. The expressions for the
eigen-energies and eigen-spinors written through the function $\varepsilon_l^-(k_\mathrm{B})$ in
Ref.~\onlinecite{Smirnov_1} are unchanged. Also the structure (that is the zero and non-zero components) of the four
dimensional eigen-spinors is the same.

The time reversal symmetry is now broken and as a result the symmetry relations between the eigen-energies and
eigen-spinors hold only if one simultaneously changes the direction of the magnetic field. For the eigen-energies we
have:
\begin{equation}
\begin{split}
&\varepsilon_{l,\eta=1}(k_\mathrm{B};H_0)=\varepsilon_{l,\eta=2}(-k_\mathrm{B};-H_0),\\
&\varepsilon_{l,\eta=3}(k_\mathrm{B};H_0)=\varepsilon_{l,\eta=4}(-k_\mathrm{B};-H_0).
\end{split}
\label{symmetry_energies}
\end{equation}
For the eigen-spinors the symmetry relations are written as:
\begin{equation}
\begin{split}
&\theta_{l,k_\mathrm{B},\eta=1}(j=\{0,1\},\sigma=\{+1,-1\};H_0)=\\
&=\theta_{l,k_\mathrm{B},\eta=2}(j=\{0,1\},\sigma=\{-1,+1\};-H_0),\\
&\theta_{l,k_\mathrm{B},\eta=3}(j=\{0,1\},\sigma=\{-1,+1\};H_0)=\\
&=\theta_{l,k_\mathrm{B},\eta=4}(j=\{0,1\},\sigma=\{+1,-1\};-H_0),
\end{split}
\label{symmetry_spinors}
\end{equation}
where it is also taken into account that the $z$-projection of the spin operator (and as a result its eigen-values)
changes its sign under the time reversal. The only non-vanishing polarization is again the one along the confinement (and
also magnetic field) direction. The symmetry relations for its components are:
\begin{equation}
\begin{split}
&P_{l,\eta=1,4}^{(z)}(k_\mathrm{B};H_0)=-P_{l,\eta=2,3}^{(z)}(-k_\mathrm{B};-H_0).
\end{split}
\label{symmetry_polarizations}
\end{equation}

Finally, we would like to note that since for the model with the first two transverse modes an operator even with respect
to $\hat{z}$ is effectively diagonal, the results obtained above remain valid with the following change. The corresponding
truly 1D problem without RSOI and transverse confinement has to be solved now not for the periodic potential $U(x)$ but
for the periodic potential $U_{\gamma,j}(x)\equiv U(x)[1+\gamma\hbar(j+1/2)/m\omega_0L^2]$. Thus the solution of that truly
1D problem acquires a dependence on the transverse mode quantum number $j$ through the periodic potential dependence on
that quantum number: $\varepsilon_l^{(0)}(k_\mathrm{B})\rightarrow\varepsilon_{\gamma,j;l}^{(0)}(k_\mathrm{B})$,
$|l,k_\mathrm{B}\rangle\rightarrow|l,k_\mathrm{B}\rangle_{\gamma,j}$. This does not change the structure (location of zero
and non-zero entries) of the resulting $4\times 4$ matrix which is thus diagonalized in the same manner as in
Ref.~\onlinecite{Smirnov_1}. We label the eigen-energies and eigen-spinors obtained from this diagonalization as
$\varepsilon_{\gamma;l,\eta}(k_\mathrm{B};H_0)$ and $\theta_{\gamma;l,k_\mathrm{B},\eta}(j,\sigma;H_0)$ to stress their
dependence on the orbit-orbit coupling strength $\gamma$. The symmetry relations
(\ref{symmetry_energies})-(\ref{symmetry_polarizations}) are, of course, unchanged.
\\
\section{Bloch states in the DVR representation}
\label{Bloch_states_and_DVR}
The scalar products $_{\gamma,j}\langle \zeta,m|l,k_\mathrm{B}\rangle_{\gamma,j}$ are nothing else than the Bloch states of
the corresponding truly 1D problem without the magnetic field and without RSOI in the representation of the coordinate
operator $\hat{x}$ operating on the subspace $\mathcal{S}\subset\mathcal{H}$, see Appendix \ref{eigen_value_x}. Thus using
the eigen-values (given in Appendix \ref{eigen_value_x}) of this coordinate operator we have:
\begin{equation}
_{\gamma,j}\langle \zeta,m|l,k_\mathrm{B}\rangle_{\gamma,j}=\mathrm{e}^{\mathrm{i}k_\mathrm{B}(mL+d_{\gamma;\zeta,j})}
u^\mathrm{DVR}_{\gamma,j;l,k_\mathrm{B}}(d_{\gamma;\zeta,j}),
\label{Bloch_state_DVR}
\end{equation}
where we denoted the Bloch amplitude with the abbreviation DVR in order to stress that it originates from the discrete
variable representation and differs from the one which originates from the continuum variable representation.

The difference of the squares of the absolute values of the hopping matrix elements,
$|\Delta^\mathrm{inter,f}_{\gamma;1,4}(m)|^2$ and $|\Delta^\mathrm{inter,f}_{\gamma;4,1}(m)|^2$, in
(\ref{stationary_averaged_spin_current_b}) can now be expressed in terms of the DVR Bloch amplitudes as
\begin{equation}
\begin{split}
&|\Delta^\mathrm{inter,f}_{\gamma;1,4}(m)|^2-|\Delta^\mathrm{inter,f}_{\gamma;4,1}(m)|^2=\\
&=-\frac{\hbar^3 k_\mathrm{so}^2\omega_0}{m}\sum_{k_\mathrm{B},k_\mathrm{B}'}\sin[(k_\mathrm{B}-k_\mathrm{B}')L]
\mathrm{Im}[F_{\gamma;k_\mathrm{B},k_\mathrm{B}'}],
\end{split}
\label{difference_of_deltas}
\end{equation}
where we have introduced a function $F_{\gamma;k_\mathrm{B},k_\mathrm{B}'}$ defined as
\begin{equation}
\begin{split}
&F_{\gamma;k_\mathrm{B},k_\mathrm{B}'}\!\!=\!
u_{\gamma,0;1,k_\mathrm{B}+k_\mathrm{so}}^\mathrm{DVR}(d_{\gamma;1,0})
u_{\gamma,1;1,k_\mathrm{B}'-k_\mathrm{so}}^\mathrm{DVR}\!(d_{\gamma;1,1})\times\\
&\times\bigl[u_{\gamma,1;1,k_\mathrm{B}-k_\mathrm{so}}^\mathrm{DVR}(d_{\gamma;1,1})
u_{\gamma,0;1,k_\mathrm{B}'+k_\mathrm{so}}^\mathrm{DVR}(d_{\gamma;1,0})\bigl]^*.
\end{split}
\label{F_function}
\end{equation}
The function $F_{\gamma;k_\mathrm{B},k_\mathrm{B}'}$ has two useful properties which directly follow from its definition
(\ref{F_function}). The first property comes from the fact that $F_{\gamma;k_\mathrm{B},k_\mathrm{B}'}$ is real if the Bloch
amplitudes are real:
\begin{equation}
\mathrm{Im}[u_{\gamma,j;1,k_\mathrm{B}}^\mathrm{DVR}(d_{\gamma;1,j})]=0\;\Rightarrow\;
\mathrm{Im}[F_{\gamma;k_\mathrm{B},k_\mathrm{B}'}]=0.
\label{real_u_real_F}
\end{equation}
The second property is that $F_{\gamma=0;k_\mathrm{B},k_\mathrm{B}'}$ is an even function in both of its arguments. Indeed, when
$\gamma=0$, we have $u_{\gamma,j;l,k_\mathrm{B}}(x)=u_{l,k_\mathrm{B}}(x)$, $d_{\gamma;\zeta,j}=d_\zeta$, that is
\begin{equation}
\begin{split}
&F_{\gamma=0;k_\mathrm{B},k_\mathrm{B}'}=
u_{1,k_\mathrm{B}+k_\mathrm{so}}^\mathrm{DVR}(d_1)u_{1,k_\mathrm{B}'-k_\mathrm{so}}^\mathrm{DVR}(d_1)\times\\
&\times\bigl[u_{1,k_\mathrm{B}-k_\mathrm{so}}^\mathrm{DVR}(d_1)
u_{1,k_\mathrm{B}'+k_\mathrm{so}}^\mathrm{DVR}(d_1)\bigl]^*.
\end{split}
\label{F_function_gamma_0}
\end{equation}
One then finds from Eq. (\ref{F_function_gamma_0}) that
$F_{\gamma=0;-k_\mathrm{B},k_\mathrm{B}'}=F_{\gamma=0;k_\mathrm{B},k_\mathrm{B}'}$ and
$F_{\gamma=0;k_\mathrm{B},-k_\mathrm{B}'}=F_{\gamma=0;k_\mathrm{B},k_\mathrm{B}'}$. As a consequence, from this property one gets
\begin{equation}
\begin{split}
&\mathrm{Im}[F_{\gamma=0;-k_\mathrm{B},k_\mathrm{B}'}]=\mathrm{Im}[F_{\gamma=0;k_\mathrm{B},k_\mathrm{B}'}],\\
&\mathrm{Im}[F_{\gamma=0;k_\mathrm{B},-k_\mathrm{B}'}]=\mathrm{Im}[F_{\gamma=0;k_\mathrm{B},k_\mathrm{B}'}],
\end{split}
\label{Im_F_even}
\end{equation}
which means that $\mathrm{Im}[F_{\gamma=0;k_\mathrm{B},k_\mathrm{B}'}]$ is even in $k_\mathrm{B}$ and $k_\mathrm{B}'$. The same is
also valid for $\mathrm{Re}[F_{\gamma=0;k_\mathrm{B},k_\mathrm{B}'}]$.


\begin{thebibliography}{44}
\expandafter\ifx\csname natexlab\endcsname\relax\def\natexlab#1{#1}\fi
\expandafter\ifx\csname bibnamefont\endcsname\relax
  \def\bibnamefont#1{#1}\fi
\expandafter\ifx\csname bibfnamefont\endcsname\relax
  \def\bibfnamefont#1{#1}\fi
\expandafter\ifx\csname citenamefont\endcsname\relax
  \def\citenamefont#1{#1}\fi
\expandafter\ifx\csname url\endcsname\relax
  \def\url#1{\texttt{#1}}\fi
\expandafter\ifx\csname urlprefix\endcsname\relax\def\urlprefix{URL }\fi
\providecommand{\bibinfo}[2]{#2}
\providecommand{\eprint}[2][]{\url{#2}}

\bibitem[{\citenamefont{Astumian and H{\"a}nggi}(2002)}]{Astumian}
\bibinfo{author}{\bibfnamefont{R.~D.} \bibnamefont{Astumian}} \bibnamefont{and}
  \bibinfo{author}{\bibfnamefont{P.}~\bibnamefont{H{\"a}nggi}},
  \bibinfo{journal}{Phys.\ Today} \textbf{\bibinfo{volume}{55}},
  \bibinfo{pages}{33} (\bibinfo{year}{2002}).

\bibitem[{\citenamefont{H{\"a}nggi and Marchesoni}()}]{Haenggi}
\bibinfo{author}{\bibfnamefont{P.}~\bibnamefont{H{\"a}nggi}} \bibnamefont{and}
  \bibinfo{author}{\bibfnamefont{F.}~\bibnamefont{Marchesoni}},
  \eprint{arXiv:0807.1283v1}.

\bibitem[{\citenamefont{Reimann et~al.}(1997)\citenamefont{Reimann, Grifoni,
  and H{\"a}nggi}}]{Reimann}
\bibinfo{author}{\bibfnamefont{P.}~\bibnamefont{Reimann}},
  \bibinfo{author}{\bibfnamefont{M.}~\bibnamefont{Grifoni}}, \bibnamefont{and}
  \bibinfo{author}{\bibfnamefont{P.}~\bibnamefont{H{\"a}nggi}},
  \bibinfo{journal}{Phys.\ Rev.\ Lett.} \textbf{\bibinfo{volume}{79}},
  \bibinfo{pages}{10} (\bibinfo{year}{1997}).

\bibitem[{\citenamefont{Grifoni et~al.}(2002)\citenamefont{Grifoni, Ferreira,
  Peguiron, and Majer}}]{Grifoni}
\bibinfo{author}{\bibfnamefont{M.}~\bibnamefont{Grifoni}},
  \bibinfo{author}{\bibfnamefont{M.~S.} \bibnamefont{Ferreira}},
  \bibinfo{author}{\bibfnamefont{J.}~\bibnamefont{Peguiron}}, \bibnamefont{and}
  \bibinfo{author}{\bibfnamefont{J.~B.} \bibnamefont{Majer}},
  \bibinfo{journal}{Phys.\ Rev.\ Lett.} \textbf{\bibinfo{volume}{89}},
  \bibinfo{pages}{146801} (\bibinfo{year}{2002}).

\bibitem[{\citenamefont{Peguiron and Grifoni}(2005)}]{Peguiron}
\bibinfo{author}{\bibfnamefont{J.}~\bibnamefont{Peguiron}} \bibnamefont{and}
  \bibinfo{author}{\bibfnamefont{M.}~\bibnamefont{Grifoni}},
  \bibinfo{journal}{Phys.\ Rev.\ E} \textbf{\bibinfo{volume}{71}},
  \bibinfo{pages}{010101(R)} (\bibinfo{year}{2005}).

\bibitem[{\citenamefont{Goychuk and H{\"a}nggi}(1998)}]{Goychuk}
\bibinfo{author}{\bibfnamefont{I.}~\bibnamefont{Goychuk}} \bibnamefont{and}
  \bibinfo{author}{\bibfnamefont{P.}~\bibnamefont{H{\"a}nggi}},
  \bibinfo{journal}{Europhys.\ Lett.} \textbf{\bibinfo{volume}{43}},
  \bibinfo{pages}{503} (\bibinfo{year}{1998}).

\bibitem[{\citenamefont{Lehmann et~al.}(2002)\citenamefont{Lehmann, Kohler,
  H{\"a}nggi, and Nitzan}}]{Lehmann}
\bibinfo{author}{\bibfnamefont{J.}~\bibnamefont{Lehmann}},
  \bibinfo{author}{\bibfnamefont{S.}~\bibnamefont{Kohler}},
  \bibinfo{author}{\bibfnamefont{P.}~\bibnamefont{H{\"a}nggi}},
  \bibnamefont{and} \bibinfo{author}{\bibfnamefont{A.}~\bibnamefont{Nitzan}},
  \bibinfo{journal}{Phys.\ Rev.\ Lett.} \textbf{\bibinfo{volume}{88}},
  \bibinfo{pages}{228305} (\bibinfo{year}{2002}).

\bibitem[{\citenamefont{Rashba}(1960)}]{Rashba}
\bibinfo{author}{\bibfnamefont{E.}~\bibnamefont{Rashba}},
  \bibinfo{journal}{Fiz.\ Tverd.\ Tela (Leningrad)}
  \textbf{\bibinfo{volume}{2}}, \bibinfo{pages}{1224} (\bibinfo{year}{1960}).

\bibitem[{\citenamefont{Murakami et~al.}(2003)\citenamefont{Murakami, Nagaosa,
  and Zhang}}]{Murakami}
\bibinfo{author}{\bibfnamefont{S.}~\bibnamefont{Murakami}},
  \bibinfo{author}{\bibfnamefont{N.}~\bibnamefont{Nagaosa}}, \bibnamefont{and}
  \bibinfo{author}{\bibfnamefont{S.-C.} \bibnamefont{Zhang}},
  \bibinfo{journal}{Science} \textbf{\bibinfo{volume}{301}},
  \bibinfo{pages}{1348} (\bibinfo{year}{2003}).

\bibitem[{\citenamefont{Sinova et~al.}(2004)\citenamefont{Sinova, Culcer, Niu,
  Sinitsyn, Jungwirth, and MacDonald}}]{Sinova}
\bibinfo{author}{\bibfnamefont{J.}~\bibnamefont{Sinova}},
  \bibinfo{author}{\bibfnamefont{D.}~\bibnamefont{Culcer}},
  \bibinfo{author}{\bibfnamefont{Q.}~\bibnamefont{Niu}},
  \bibinfo{author}{\bibfnamefont{N.~A.} \bibnamefont{Sinitsyn}},
  \bibinfo{author}{\bibfnamefont{T.}~\bibnamefont{Jungwirth}},
  \bibnamefont{and} \bibinfo{author}{\bibfnamefont{A.~H.}
  \bibnamefont{MacDonald}}, \bibinfo{journal}{Phys.\ Rev.\ Lett.}
  \textbf{\bibinfo{volume}{92}}, \bibinfo{pages}{126603}
  (\bibinfo{year}{2004}).

\bibitem[{\citenamefont{Wunderlich et~al.}(2005)\citenamefont{Wunderlich,
  Kaestner, Sinova, and Jungwirth}}]{Wunderlich}
\bibinfo{author}{\bibfnamefont{J.}~\bibnamefont{Wunderlich}},
  \bibinfo{author}{\bibfnamefont{B.}~\bibnamefont{Kaestner}},
  \bibinfo{author}{\bibfnamefont{J.}~\bibnamefont{Sinova}}, \bibnamefont{and}
  \bibinfo{author}{\bibfnamefont{T.}~\bibnamefont{Jungwirth}},
  \bibinfo{journal}{Phys.\ Rev.\ Lett.} \textbf{\bibinfo{volume}{94}},
  \bibinfo{pages}{047204} (\bibinfo{year}{2005}).

\bibitem[{\citenamefont{Kato et~al.}(2004{\natexlab{a}})\citenamefont{Kato,
  Myers, Gossard, and Awschalom}}]{Kato}
\bibinfo{author}{\bibfnamefont{Y.~K.} \bibnamefont{Kato}},
  \bibinfo{author}{\bibfnamefont{R.~C.} \bibnamefont{Myers}},
  \bibinfo{author}{\bibfnamefont{A.~C.} \bibnamefont{Gossard}},
  \bibnamefont{and} \bibinfo{author}{\bibfnamefont{D.~D.}
  \bibnamefont{Awschalom}}, \bibinfo{journal}{Science}
  \textbf{\bibinfo{volume}{306}}, \bibinfo{pages}{1910}
  (\bibinfo{year}{2004}{\natexlab{a}}).

\bibitem[{\citenamefont{Hankiewicz et~al.}(2004)\citenamefont{Hankiewicz,
  Molenkamp, Jungwirth, and Sinova}}]{Hankiewicz}
\bibinfo{author}{\bibfnamefont{E.~M.} \bibnamefont{Hankiewicz}},
  \bibinfo{author}{\bibfnamefont{L.~W.} \bibnamefont{Molenkamp}},
  \bibinfo{author}{\bibfnamefont{T.}~\bibnamefont{Jungwirth}},
  \bibnamefont{and} \bibinfo{author}{\bibfnamefont{J.}~\bibnamefont{Sinova}},
  \bibinfo{journal}{Phys.\ Rev.\ B} \textbf{\bibinfo{volume}{70}},
  \bibinfo{pages}{241301(R)} (\bibinfo{year}{2004}).

\bibitem[{\citenamefont{Hankiewicz et~al.}(2005)\citenamefont{Hankiewicz, Li,
  Jungwirth, Niu, Shen, and Sinova}}]{Hankiewicz_1}
\bibinfo{author}{\bibfnamefont{E.~M.} \bibnamefont{Hankiewicz}},
  \bibinfo{author}{\bibfnamefont{J.}~\bibnamefont{Li}},
  \bibinfo{author}{\bibfnamefont{T.}~\bibnamefont{Jungwirth}},
  \bibinfo{author}{\bibfnamefont{Q.}~\bibnamefont{Niu}},
  \bibinfo{author}{\bibfnamefont{S.-Q.} \bibnamefont{Shen}}, \bibnamefont{and}
  \bibinfo{author}{\bibfnamefont{J.}~\bibnamefont{Sinova}},
  \bibinfo{journal}{Phys.\ Rev.\ B} \textbf{\bibinfo{volume}{72}},
  \bibinfo{pages}{155305} (\bibinfo{year}{2005}).

\bibitem[{\citenamefont{Valenzuela and Tinkham}(2006)}]{Valenzuela}
\bibinfo{author}{\bibfnamefont{S.~O.} \bibnamefont{Valenzuela}}
  \bibnamefont{and} \bibinfo{author}{\bibfnamefont{M.}~\bibnamefont{Tinkham}},
  \bibinfo{journal}{Nature (London)} \textbf{\bibinfo{volume}{442}},
  \bibinfo{pages}{176} (\bibinfo{year}{2006}).

\bibitem[{\citenamefont{Bhat et~al.}(2005)\citenamefont{Bhat, Nastos, Najmaie,
  and Sipe}}]{Bhat}
\bibinfo{author}{\bibfnamefont{R.~D.~R.} \bibnamefont{Bhat}},
  \bibinfo{author}{\bibfnamefont{F.}~\bibnamefont{Nastos}},
  \bibinfo{author}{\bibfnamefont{A.}~\bibnamefont{Najmaie}}, \bibnamefont{and}
  \bibinfo{author}{\bibfnamefont{J.~E.} \bibnamefont{Sipe}},
  \bibinfo{journal}{Phys.\ Rev.\ Lett.} \textbf{\bibinfo{volume}{94}},
  \bibinfo{pages}{096603} (\bibinfo{year}{2005}).

\bibitem[{\citenamefont{Li et~al.}(2006)\citenamefont{Li, Dai, Shen, and
  Zhang}}]{Li}
\bibinfo{author}{\bibfnamefont{J.}~\bibnamefont{Li}},
  \bibinfo{author}{\bibfnamefont{X.}~\bibnamefont{Dai}},
  \bibinfo{author}{\bibfnamefont{S.-Q.} \bibnamefont{Shen}}, \bibnamefont{and}
  \bibinfo{author}{\bibfnamefont{F.-C.} \bibnamefont{Zhang}},
  \bibinfo{journal}{Appl.\ Phys.\ Lett.} \textbf{\bibinfo{volume}{88}},
  \bibinfo{pages}{162105} (\bibinfo{year}{2006}).

\bibitem[{\citenamefont{Zhou and Shen}(2007)}]{Zhou}
\bibinfo{author}{\bibfnamefont{B.}~\bibnamefont{Zhou}} \bibnamefont{and}
  \bibinfo{author}{\bibfnamefont{S.-Q.} \bibnamefont{Shen}},
  \bibinfo{journal}{Phys.\ Rev.\ B} \textbf{\bibinfo{volume}{75}},
  \bibinfo{pages}{045339} (\bibinfo{year}{2007}).

\bibitem[{\citenamefont{Altshuler and Glazman}(1999)}]{Altshuler}
\bibinfo{author}{\bibfnamefont{B.~L.} \bibnamefont{Altshuler}}
  \bibnamefont{and} \bibinfo{author}{\bibfnamefont{L.~I.}
  \bibnamefont{Glazman}}, \bibinfo{journal}{Science}
  \textbf{\bibinfo{volume}{283}}, \bibinfo{pages}{1864} (\bibinfo{year}{1999}).

\bibitem[{\citenamefont{Sharma and Chamon}(2001)}]{Sharma}
\bibinfo{author}{\bibfnamefont{P.}~\bibnamefont{Sharma}} \bibnamefont{and}
  \bibinfo{author}{\bibfnamefont{C.}~\bibnamefont{Chamon}},
  \bibinfo{journal}{Phys.\ Rev.\ Lett.} \textbf{\bibinfo{volume}{87}},
  \bibinfo{pages}{096401} (\bibinfo{year}{2001}).

\bibitem[{\citenamefont{Benjamin and Benjamin}(2004)}]{Benjamin}
\bibinfo{author}{\bibfnamefont{R.}~\bibnamefont{Benjamin}} \bibnamefont{and}
  \bibinfo{author}{\bibfnamefont{C.}~\bibnamefont{Benjamin}},
  \bibinfo{journal}{Phys.\ Rev.\ B} \textbf{\bibinfo{volume}{69}},
  \bibinfo{pages}{085318} (\bibinfo{year}{2004}).

\bibitem[{\citenamefont{Wei et~al.}(2004)\citenamefont{Wei, Wan, Wang, and
  Wang}}]{Wei}
\bibinfo{author}{\bibfnamefont{Y.}~\bibnamefont{Wei}},
  \bibinfo{author}{\bibfnamefont{L.}~\bibnamefont{Wan}},
  \bibinfo{author}{\bibfnamefont{B.}~\bibnamefont{Wang}}, \bibnamefont{and}
  \bibinfo{author}{\bibfnamefont{J.}~\bibnamefont{Wang}},
  \bibinfo{journal}{Phys.\ Rev.\ B} \textbf{\bibinfo{volume}{70}},
  \bibinfo{pages}{045418} (\bibinfo{year}{2004}).

\bibitem[{\citenamefont{Watson et~al.}(2003)\citenamefont{Watson, Potok,
  Marcus, and Umansky}}]{Watson}
\bibinfo{author}{\bibfnamefont{S.~K.} \bibnamefont{Watson}},
  \bibinfo{author}{\bibfnamefont{R.~M.} \bibnamefont{Potok}},
  \bibinfo{author}{\bibfnamefont{C.~M.} \bibnamefont{Marcus}},
  \bibnamefont{and} \bibinfo{author}{\bibfnamefont{V.}~\bibnamefont{Umansky}},
  \bibinfo{journal}{Phys.\ Rev.\ Lett.} \textbf{\bibinfo{volume}{91}},
  \bibinfo{pages}{258301} (\bibinfo{year}{2003}).

\bibitem[{\citenamefont{Scheid et~al.}(2007{\natexlab{a}})\citenamefont{Scheid,
  Pfund, Bercioux, and Richter}}]{Scheid}
\bibinfo{author}{\bibfnamefont{M.}~\bibnamefont{Scheid}},
  \bibinfo{author}{\bibfnamefont{A.}~\bibnamefont{Pfund}},
  \bibinfo{author}{\bibfnamefont{D.}~\bibnamefont{Bercioux}}, \bibnamefont{and}
  \bibinfo{author}{\bibfnamefont{K.}~\bibnamefont{Richter}},
  \bibinfo{journal}{Phys.\ Rev.\ B} \textbf{\bibinfo{volume}{76}},
  \bibinfo{pages}{195303} (\bibinfo{year}{2007}{\natexlab{a}}).

\bibitem[{\citenamefont{Smirnov et~al.}(2008)\citenamefont{Smirnov, Bercioux,
  Grifoni, and Richter}}]{Smirnov}
\bibinfo{author}{\bibfnamefont{S.}~\bibnamefont{Smirnov}},
  \bibinfo{author}{\bibfnamefont{D.}~\bibnamefont{Bercioux}},
  \bibinfo{author}{\bibfnamefont{M.}~\bibnamefont{Grifoni}}, \bibnamefont{and}
  \bibinfo{author}{\bibfnamefont{K.}~\bibnamefont{Richter}},
  \bibinfo{journal}{Phys.\ Rev.\ Lett.} \textbf{\bibinfo{volume}{100}},
  \bibinfo{pages}{230601} (\bibinfo{year}{2008}).

\bibitem[{\citenamefont{Scheid et~al.}(2007{\natexlab{b}})\citenamefont{Scheid,
  Bercioux, and Richter}}]{Scheid_1}
\bibinfo{author}{\bibfnamefont{M.}~\bibnamefont{Scheid}},
  \bibinfo{author}{\bibfnamefont{D.}~\bibnamefont{Bercioux}}, \bibnamefont{and}
  \bibinfo{author}{\bibfnamefont{K.}~\bibnamefont{Richter}},
  \bibinfo{journal}{New\ J.\ Phys.} \textbf{\bibinfo{volume}{9}},
  \bibinfo{pages}{401} (\bibinfo{year}{2007}{\natexlab{b}}).

\bibitem[{\citenamefont{Caldeira and Leggett}(1981)}]{Caldeira}
\bibinfo{author}{\bibfnamefont{A.~O.} \bibnamefont{Caldeira}} \bibnamefont{and}
  \bibinfo{author}{\bibfnamefont{A.~J.} \bibnamefont{Leggett}},
  \bibinfo{journal}{Phys.\ Rev.\ Lett.} \textbf{\bibinfo{volume}{46}},
  \bibinfo{pages}{211} (\bibinfo{year}{1981}).

\bibitem[{\citenamefont{Caldeira and Leggett}(1983)}]{Caldeira_1}
\bibinfo{author}{\bibfnamefont{A.~O.} \bibnamefont{Caldeira}} \bibnamefont{and}
  \bibinfo{author}{\bibfnamefont{A.~J.} \bibnamefont{Leggett}},
  \bibinfo{journal}{Ann.\ Phys.} \textbf{\bibinfo{volume}{149}},
  \bibinfo{pages}{374} (\bibinfo{year}{1983}).

\bibitem[{\citenamefont{Shi et~al.}(2006)\citenamefont{Shi, Zhang, Xiao, and
  Niu}}]{Shi}
\bibinfo{author}{\bibfnamefont{J.}~\bibnamefont{Shi}},
  \bibinfo{author}{\bibfnamefont{P.}~\bibnamefont{Zhang}},
  \bibinfo{author}{\bibfnamefont{D.}~\bibnamefont{Xiao}}, \bibnamefont{and}
  \bibinfo{author}{\bibfnamefont{Q.}~\bibnamefont{Niu}},
  \bibinfo{journal}{Phys.\ Rev.\ Lett.} \textbf{\bibinfo{volume}{96}},
  \bibinfo{pages}{076604} (\bibinfo{year}{2006}).

\bibitem[{\citenamefont{Zhang et~al.}(2008)\citenamefont{Zhang, Wang, Shi,
  Xiao, and Niu}}]{Zhang}
\bibinfo{author}{\bibfnamefont{P.}~\bibnamefont{Zhang}},
  \bibinfo{author}{\bibfnamefont{Z.}~\bibnamefont{Wang}},
  \bibinfo{author}{\bibfnamefont{J.}~\bibnamefont{Shi}},
  \bibinfo{author}{\bibfnamefont{D.}~\bibnamefont{Xiao}}, \bibnamefont{and}
  \bibinfo{author}{\bibfnamefont{Q.}~\bibnamefont{Niu}},
  \bibinfo{journal}{Phys.\ Rev.\ B} \textbf{\bibinfo{volume}{77}},
  \bibinfo{pages}{075304} (\bibinfo{year}{2008}).

\bibitem[{\citenamefont{Sih et~al.}(2005)\citenamefont{Sih, Myers, Kato, Lau,
  Gossard, and Awschalom}}]{Sih}
\bibinfo{author}{\bibfnamefont{V.}~\bibnamefont{Sih}},
  \bibinfo{author}{\bibfnamefont{R.~C.} \bibnamefont{Myers}},
  \bibinfo{author}{\bibfnamefont{Y.~K.} \bibnamefont{Kato}},
  \bibinfo{author}{\bibfnamefont{W.~H.} \bibnamefont{Lau}},
  \bibinfo{author}{\bibfnamefont{A.~C.} \bibnamefont{Gossard}},
  \bibnamefont{and} \bibinfo{author}{\bibfnamefont{D.~D.}
  \bibnamefont{Awschalom}}, \bibinfo{journal}{Nat.\ Phys.}
  \textbf{\bibinfo{volume}{1}}, \bibinfo{pages}{31} (\bibinfo{year}{2005}).

\bibitem[{\citenamefont{Kato et~al.}(2004{\natexlab{b}})\citenamefont{Kato,
  Myers, Gossard, and Awschalom}}]{Kato_1}
\bibinfo{author}{\bibfnamefont{Y.~K.} \bibnamefont{Kato}},
  \bibinfo{author}{\bibfnamefont{R.~C.} \bibnamefont{Myers}},
  \bibinfo{author}{\bibfnamefont{A.~C.} \bibnamefont{Gossard}},
  \bibnamefont{and} \bibinfo{author}{\bibfnamefont{D.~D.}
  \bibnamefont{Awschalom}}, \bibinfo{journal}{Phys.\ Rev.\ Lett.}
  \textbf{\bibinfo{volume}{93}}, \bibinfo{pages}{176601}
  (\bibinfo{year}{2004}{\natexlab{b}}).

\bibitem[{\citenamefont{Kotissek et~al.}(2007)\citenamefont{Kotissek, Bailleul,
  Sperl, Spitzer, Schuh, Wegscheider, Back, and Bayreuther}}]{Kotissek}
\bibinfo{author}{\bibfnamefont{P.}~\bibnamefont{Kotissek}},
  \bibinfo{author}{\bibfnamefont{M.}~\bibnamefont{Bailleul}},
  \bibinfo{author}{\bibfnamefont{M.}~\bibnamefont{Sperl}},
  \bibinfo{author}{\bibfnamefont{A.}~\bibnamefont{Spitzer}},
  \bibinfo{author}{\bibfnamefont{D.}~\bibnamefont{Schuh}},
  \bibinfo{author}{\bibfnamefont{W.}~\bibnamefont{Wegscheider}},
  \bibinfo{author}{\bibfnamefont{C.~H.} \bibnamefont{Back}}, \bibnamefont{and}
  \bibinfo{author}{\bibfnamefont{G.}~\bibnamefont{Bayreuther}},
  \bibinfo{journal}{Nat.\ Phys.} \textbf{\bibinfo{volume}{3}},
  \bibinfo{pages}{872} (\bibinfo{year}{2007}).

\bibitem[{\citenamefont{Smirnov et~al.}(2007)\citenamefont{Smirnov, Bercioux,
  and Grifoni}}]{Smirnov_1}
\bibinfo{author}{\bibfnamefont{S.}~\bibnamefont{Smirnov}},
  \bibinfo{author}{\bibfnamefont{D.}~\bibnamefont{Bercioux}}, \bibnamefont{and}
  \bibinfo{author}{\bibfnamefont{M.}~\bibnamefont{Grifoni}},
  \bibinfo{journal}{EPL} \textbf{\bibinfo{volume}{80}}, \bibinfo{pages}{27003}
  (\bibinfo{year}{2007}).

\bibitem[{\citenamefont{Harris et~al.}(1965)\citenamefont{Harris, Engerholm,
  and Gwinn}}]{Harris}
\bibinfo{author}{\bibfnamefont{D.~O.} \bibnamefont{Harris}},
  \bibinfo{author}{\bibfnamefont{G.~G.} \bibnamefont{Engerholm}},
  \bibnamefont{and} \bibinfo{author}{\bibfnamefont{W.~D.} \bibnamefont{Gwinn}},
  \bibinfo{journal}{J.\ Chem.\ Phys.} \textbf{\bibinfo{volume}{43}},
  \bibinfo{pages}{1515} (\bibinfo{year}{1965}).

\bibitem[{\citenamefont{Grifoni and H{\"a}nggi}(1998)}]{Grifoni_1}
\bibinfo{author}{\bibfnamefont{M.}~\bibnamefont{Grifoni}} \bibnamefont{and}
  \bibinfo{author}{\bibfnamefont{P.}~\bibnamefont{H{\"a}nggi}},
  \bibinfo{journal}{Phys.\ Rep.} \textbf{\bibinfo{volume}{304}},
  \bibinfo{pages}{229} (\bibinfo{year}{1998}).

\bibitem[{\citenamefont{Weiss}(1999)}]{Weiss}
\bibinfo{author}{\bibfnamefont{U.}~\bibnamefont{Weiss}},
  \emph{\bibinfo{title}{Quantum Dissipative Systems}}
  (\bibinfo{publisher}{World Scientific, Singapore}, \bibinfo{year}{1999}),
  \bibinfo{edition}{2nd} ed.

\bibitem[{\citenamefont{Perroni et~al.}(2007)\citenamefont{Perroni, Bercioux,
  Ramaglia, and Cataudella}}]{Perroni}
\bibinfo{author}{\bibfnamefont{C.~A.} \bibnamefont{Perroni}},
  \bibinfo{author}{\bibfnamefont{D.}~\bibnamefont{Bercioux}},
  \bibinfo{author}{\bibfnamefont{V.~M.} \bibnamefont{Ramaglia}},
  \bibnamefont{and}
  \bibinfo{author}{\bibfnamefont{V.}~\bibnamefont{Cataudella}},
  \bibinfo{journal}{J.\ Phys.\: Condens.\ Matter}
  \textbf{\bibinfo{volume}{19}}, \bibinfo{pages}{186227}
  (\bibinfo{year}{2007}).

\bibitem[{\citenamefont{Hartmann et~al.}(1997)\citenamefont{Hartmann, Grifoni,
  and H{\"a}nggi}}]{Hartmann}
\bibinfo{author}{\bibfnamefont{L.}~\bibnamefont{Hartmann}},
  \bibinfo{author}{\bibfnamefont{M.}~\bibnamefont{Grifoni}}, \bibnamefont{and}
  \bibinfo{author}{\bibfnamefont{P.}~\bibnamefont{H{\"a}nggi}},
  \bibinfo{journal}{Europhys.\ Lett.} \textbf{\bibinfo{volume}{38}},
  \bibinfo{pages}{497} (\bibinfo{year}{1997}).

\bibitem[{\citenamefont{Rashba}(2003)}]{Rashba_1}
\bibinfo{author}{\bibfnamefont{E.~I.} \bibnamefont{Rashba}},
  \bibinfo{journal}{Phys.\ Rev.\ B} \textbf{\bibinfo{volume}{68}},
  \bibinfo{pages}{241315(R)} (\bibinfo{year}{2003}).

\bibitem[{\citenamefont{Sch{\"a}pers et~al.}(2004)\citenamefont{Sch{\"a}pers,
  Knobbe, and Guzenko}}]{Schaepers}
\bibinfo{author}{\bibfnamefont{T.}~\bibnamefont{Sch{\"a}pers}},
  \bibinfo{author}{\bibfnamefont{J.}~\bibnamefont{Knobbe}}, \bibnamefont{and}
  \bibinfo{author}{\bibfnamefont{V.~A.} \bibnamefont{Guzenko}},
  \bibinfo{journal}{Phys.\ Rev.\ B} \textbf{\bibinfo{volume}{69}},
  \bibinfo{pages}{235323} (\bibinfo{year}{2004}).

\bibitem[{\citenamefont{Madelung}(2003)}]{Madelung}
\bibinfo{author}{\bibfnamefont{O.}~\bibnamefont{Madelung}},
  \emph{\bibinfo{title}{Semiconductors: Data Handbook}}
  (\bibinfo{publisher}{Springer}, \bibinfo{address}{Berlin},
  \bibinfo{year}{2003}).

\bibitem[{\citenamefont{Steinshnider et~al.}(2000)\citenamefont{Steinshnider,
  Harper, Weimer, Lin, \text{S.S.} Pei, and \text{D.H.} Chow}}]{Steinshnider}
\bibinfo{author}{\bibfnamefont{J.}~\bibnamefont{Steinshnider}},
  \bibinfo{author}{\bibfnamefont{J.}~\bibnamefont{Harper}},
  \bibinfo{author}{\bibfnamefont{M.}~\bibnamefont{Weimer}},
  \bibinfo{author}{\bibfnamefont{C.-H.} \bibnamefont{Lin}},
  \bibinfo{author}{\bibnamefont{\text{S.S.} Pei}}, \bibnamefont{and}
  \bibinfo{author}{\bibnamefont{\text{D.H.} Chow}}, \bibinfo{journal}{Phys.\
  Rev.\ Lett.} \textbf{\bibinfo{volume}{85}}, \bibinfo{pages}{4562}
  (\bibinfo{year}{2000}).

\bibitem[{\citenamefont{\text{L.V.} Keldysh}(1965)}]{Keldysh}
\bibinfo{author}{\bibnamefont{\text{L.V.} Keldysh}}, \bibinfo{journal}{Sov.
  Phys. JETP} \textbf{\bibinfo{volume}{20}}, \bibinfo{pages}{1018}
  (\bibinfo{year}{1965}).

\end{thebibliography}
\end{document}